\documentclass[12pt]{article}
\usepackage{geometry}
\usepackage{graphicx}
\usepackage{amsmath}
\usepackage{nccmath}
\usepackage{amssymb}
\usepackage{caption}
\usepackage{subcaption}
\usepackage{cite}
\usepackage{hyperref}
\usepackage{enumitem} 
\usepackage{titlesec}
\titleformat{\section}{\large\bfseries}{\thesection}{1em}{}
\titleformat{\subsection}{\normalsize\bfseries}{\thesubsection}{1em}{}
\providecommand{\keywords}[1]{\small \textbf{\textit{Keywords---}} #1}
\makeatletter
\def\maketitle{
	\begin{center}
		{\LARGE \bfseries \@title}\par
		\vskip 0.5em  
		{\large \@author}\par
		\vskip 0.2em 
	\end{center}
	\thispagestyle{plain}  
	\setcounter{footnote}{0} 
}
\makeatother

\title{\large \textbf{Exact solution of Schrödinger equation for the complex Morse potential to investigate physical systems with position-dependent complex mass}}
\author{\normalsize Partha Sarathi\textsuperscript{1} 
	\hspace{-0.4em} \normalsize and
	Bhaskar Singh Rawat\textsuperscript{2} \\
	\small \textsuperscript{1}Department of Physics, Maharaja Agrasen College, University of Delhi, Vasundhara Enclave, Delhi-110096, India \\
	\textsuperscript{2}Department of Physics, Hemvati Nandan Bahuguna Garhwal University, Srinagar-246174, Uttarakhand, India
}

\date{} 

\begin{document}
	
	\maketitle
	\vspace{-0.5em} 
	
	\footnotetext[1]{\texttt{parthasarthi@mac.du.ac.in}}
	\footnotetext[2]{\texttt{bhaskarsinghrawat20@gmail.com}}
	
	\begin{abstract}
	This paper presents the exact ground state solution for a diatomic particle system with position-dependent complex mass under action of a complex Morse potential in the quantum domain. By solving the position-dependent Schrödinger equation in extended complex phase space without assuming a specific mass profile, we derive both the eigenfunctions and corresponding eigenenergies using the analyticity conditions of the eigenfunctions. A key focus is placed on addressing the challenge of normalization inherent in non-Hermitian Hamiltonians. To overcome the limitations of conventional normalization methods in systems with complex potentials and spatially varying mass, we propose a modified normalization approach based on a two-dimensional integral over phase space. The results reveal that, under certain parameter constraints, real energy spectra can arise in non-Hermitian settings, supported by normalized and physically meaningful eigenfunctions. Probability density plots validate the existence of stable, localized bound states, maintaining essential characteristics of the traditional Morse potential. Moreover, the model offers potential applications in high-energy and cosmological physics, particularly in the quantum
	description of exotic systems like dark matter.
	\end{abstract} 
	\keywords{exact solution, real eigenvalues, complex Morse potential, position-dependent mass, complex mass, bound states, dark matter}
	
	\noindent\rule{\linewidth}{0.4pt}
	
	\section{Introduction} \label{sec1}
	
	The study of non-Hermitian operators, particularly in the realm of finding the solution of Schrödinger
	equation with complex potentials, has gained significant attention in last three decades due to its
	applications in open quantum systems, optical systems, and resonance phenomena. Unlike self-
	adjoint operators, non-Hermitian operators may possess complex eigenvalues, non-orthogonal
	eigenfunctions, and lack a complete set of eigenvectors. Foundational contributions by T. Kato \cite{kato2013perturbation}
	laid the groundwork for understanding the spectral properties of non-self-adjoint operators,
	including spectral instabilities and the emergence of Jordan blocks. E. B. Davies \cite{davies2007linear} explored
	pseudospectra and spectral pollution, emphasizing the physical and numerical consequences of non-
	normal operators. The framework of PT-symmetric and pseudo-Hermitian quantum mechanics,
	developed by C. M. Bender \cite{bender1998real} and A. Mostafazadeh \cite{mostafazadeh2002pseudo} respectively, showed that certain non-
	Hermitian Hamiltonians can still have entirely real spectra under symmetry constraints. Rigged
	Hilbert spaces and Hardy spaces have also been employed to describe decaying states and
	resonances rigorously, particularly in complex scaling and scattering theory \cite{vilenkin1964generalized}. These developments
	reveal that the spectral analysis of non-Hermitian systems requires tools beyond the standard
	Hilbert space formalism and often demands functional analytic techniques tailored to generalized or
	dissipative operators \cite{reed1978iv,strauss2007partial}. Our work aims to position itself within this broader context by
	incorporating these insights where applicable.
	
	Systems with position-dependent mass have garnered significant interest from
	researchers in recent decades due to their diverse applications across fundamental
	areas of physics. These systems can be analyzed from both classical and quantum
	perspectives. In the quantum domain, recent research has focused on obtaining the
	solution of the Schrödinger equation with a position-dependent mass, exploring
	various studies in this area \cite{dekar1998exactly, yu2004exactly, alhaidari2002solutions, mirabotalebi2006conformal}. The concept of position-dependent mass has been
	applied in numerous fields, including semiconductors \cite{von1983position,von1985position, zhu1983interface, weishbuch1993quantum, el2020generalized, el2020new, bastard1988wave}, quantum wells and
	quantum dots \cite{harrison2016quantum, wibawa2024effect,li1993band, rajashabala2006effective, peter2009effect, khordad2011effect, condmat8040086}, quantum liquids \cite{de1994effective}, and impurities in crystals \cite{geller1993quantum}, graded mixed semiconductors \cite{gora1969theory}, abrupt heterostructures \cite{einevoll1988effective, morrow1984model} and in lattice and superlattice
	systems \cite{bastard1981superlattice, el2021dynamics}. Additionally, they play a role in investigating inversion potentials in
	molecules using density functional theory \cite{aquino1998inversion}, many-body systems \cite{bencheikh2004extended}, the energy
	spectrum and transition rates of atomic nuclei \cite{alimohammadi2017investigation}, magnetic monopoles \cite{de2019exact}, as well
	as classical and quantum nonlinear oscillators \cite{mathews1974unique, carinena2008quantum, ballesteros2011two, mustafa2021isochronous} and curved spaces \cite{quesne2004deformed}.
	
	Further, the investigation of various potentials in the quantum domain with position-dependent mass has increasingly captured the attention of theoretical physicists.
	Recent studies have explored position-dependent mass carriers in different potential
	scenarios, such as a 3-dimensional cylindrical wire with a hyperbolic potential \cite{christiansen2023three}, an
	oscillator-shaped quantum well \cite{quesne2023rational}, a Pöschl-Teller-type potential \cite{eyube2023energy}, and a Morse
	potential \cite{Bagchi_2023}.
	
	The Morse potential has been effectively used to describe the vibrational and
	rotational behaviour of diatomic molecules \cite{morse1929diatomic}. Although several studies \cite{chen2004exact, plastino1999supersymmetric, bagchi2006morse, rajbongshi2015generation, ikhdair2012effective, arda2011bound, ovando2013three, moya2014squeeze}
	have investigated solutions to the Schrödinger equation with a position-dependent
	mass for the Morse potential, these approaches typically involve approximation
	techniques. Such methods often rely on assumptions about the mass function or other
	parameters of the potential, rather than providing exact solutions to the Schrödinger
	equation.
	
	Although several efforts have been made to find the solution of the Schrödinger
	equation for the potential dependent masses, little effort has been allotted to obtain the
	exact solution for such potentials in the quantum domain. Some efforts have been
	made to develop a quantum kinetic energy operator within the Schrödinger equation to
	create a new class of exactly solvable models featuring a position-dependent mass
	\cite{ganguly2006study}, they primarily focus on the real potentials \cite{dhahbi2019new, ikot2024exact}. The study of complex
	potentials has gathered interest due to their novel properties and their probable
	applications in the realm of various branches of science to understand the behaviour of
	complex physical systems. An extended complex phase space formulation of
	Schrödinger quantum mechanics \cite{kaushal2002quantum, kaushal2003quantum, parashar2004complex, chand2007solution, singh2009solution, singh2013closed} has been employed to study the complex
	potentials and gain insight into understanding the quantum dynamics of Non-Hermitian Hamiltonian systems (NHH) systems. The same formulation was used to
	establish that it is indeed feasible for negative masses for various physical systems to
	be bound together by complex Morse potential. It was established that atoms with
	negative masses may bind together under the action of complex Morse-like potentials and form molecular structures \cite{sarathi2021application}. Furthermore, in the domain of relativistic quantum mechanics, Özlem et.al \cite{yecsiltacs2023dirac} arrived at the solution of the Dirac
	equation in the frame of position-dependent mass where the eigenvalues of the Dirac
	operator for the complex Morse and trigonometric complex Scarf-II potentials SL(2,C) using Lie algebras and the supersymmetric quantum mechanical approaches were
	obtained.
	
	The study of dark matter in the realm of Cosmology and Quantum Mechanics poses
	serious challenges as the only concrete evidence for dark matter (DM) stems from its
	gravitational interactions. Determining the nature of DM is one of the principal
	challenges of modern experimental physics. The theoretical investigations of position
	dependent masses in the quantum domain have opened new vistas for arriving at a
	plausible theory to explain the characteristics of the DM. The system of stability
	equations for galactic halos is under-determined in most of the models of dark matter
	(DM) which has been conventionally resolved by taking the temperature as a constant,
	and the chemical potential and the mass density as position-dependent
	variables\cite{borzou2021estimation}.The position-dependent probability distribution function (PDF) of the
	smoothed matter field as a cosmological observable was introduced to highlight
	position-dependent PDF and its distinct dependence on cosmological parameters\cite{jamieson2020position}.
	By considering the diffusion coefficient as increasing exponentially with the distance
	from the disk, it was established that the antiproton flux predicted deviates from the
	conventional calculation for the same dark matter parameters by up to about 25\% \cite{perelstein2011antiprotons}.
	A study of the position-dependent Voronoi density PDF, measured within finite
	subregions of the Universe, using separate universe simulations, demonstrate that the
	spatial variation of the position-dependent PDF is due to large-scale density fluctuations \cite{PhysRevD.103.103522}. Brax et.al \cite{PhysRevD.105.103015} explored alternatives to the collisionless cold dark
	matter paradigm by proposing a simple two-component dark matter model that
	features derivative conformal interactions between the two DM species. In their
	model, they considered temperature and the chemical potential as position-dependent
	variables. Other studies have forwarded arguments in favor of the mass-dependent
	maximal acceleration (MDMA) by proposing a hypothesis that there exists a maximal
	force with the numerical value equal to the inverse Einstein's gravitation constant \cite{tomilchik2005conformallyflatmetricpositiondependent}.
	Attempt has been made recently to prove that the molecules in the dark matter which
	are assumed to be negative complex masses can bind together under the action of
	complex Morse-like potentials and form molecular structures\cite{sarathi2021application}.
	
	With the explanation of dark matter still an open problem, an attempt is made in this
	paper to study the position-dependent complex mass (PDCM) under action of complex morse potential in the
	quantum domain. Further, the results obtained are analyzed to forward a quantum
	mechanical explanation of the composition of dark matter. We propose that the dark
	matter can be a PDCM under the action of complex
	Morse potential. Thus, we find the exact solution of the  Schrödinger equation for the
	one-dimensional complex Morse potential with PDCM,
	
	$$V(x) = V_0\left[e^{-2ax}-2e^{-ax}\right].$$
	
	where $V_0$ is the well depth, $x$ is the internuclear distance (bond length), $x>0$ and $a$ is related
	to the vibrational constant $\mu$ as
	
	$$a = \pi \mu \sqrt{\frac{2 M}{V_0}}.$$
	
	The reduced mass $M$ of the system is considered a position-dependent mass and is a
	function of the internuclear distance $x$ which is taken as complex. In general, the
	parameters $V_0$ and $a$ are also considered complex as the objective of the study is to
	investigate the bonding between two particles of PDCM.
	
	The arrangement of the paper is as follows:
	In Section \ref{sec2}, a general formulation for the solution of the  Schrödinger equation for a
	general class of complex potential is enumerated. In Section \ref{sec3}, the exact solution of
	the  Schrödinger equation is obtained for the general class of one-dimensional
	complex Morse potential with PDCM and its eigenvalues
	and eigenfunction are computed. Further the normalized eigenfunction and probablity density is derived in Section \ref{sec4}. The exact solution of normalized eigenfunction and eigenvalues are implimented in Section \ref{sec5} for the general case of PDCM. The admissibility of real eigenvalues is discussed in
	Section \ref{sec6}. Lastly, a general discussion on the results and applications of such studies is carried forward in Section
	\ref{sec7}.
	\section{Schrödinger equation for PDM and its general solution} \label{sec2}
	
	This paper aims to investigate the quantum behaviour of a diatomic particle system with
	a position-dependent complex mass under the influence of a complex Morse potential.
	By solving the position-dependent Schrödinger equation (PDMSE) for the above-
	mentioned potential in extended complex phase space, we obtain exact ground-state
	eigenfunctions and energies. Unlike previous attempts in the literature to solve the
	Schrodinger equation for position-dependent real Morse potentials, no specific mass
	profile is assumed. The approach relies on analyticity conditions to ensure physical
	solutions that can also admit real energy spectra under certain constraining relations
	amongst the potential parameters. This framework reveals that position-dependent
	complex mass systems admit normalized eigenfunctions and can support bound states,
	expanding the scope of quantum models that can be applied in various physical
	systems like quantum dots, quantum wells, dark matter etc.
	
	The solution to the position-dependent mass  Schrödinger equation (PDMSE) can be expressed by reformulating the position $x$ and momentum $p$ in complex phase space, which is increasingly advocated in recent years due to its potentially more rigorous mathematical foundation, as given by \cite{XAVIERJR1996458}
	
	\begin{equation}
		x = x_1 + ip_2, \qquad p = p_1 + ix_2. \label{eq1}
	\end{equation}
	where the imaginary parts $x_2$ and $p_2$ introduced in the variables $p$ and $x$ turn out to be canonical pairs of $p_1$ and $x_1$ respectively. The complexification of phase space enumerated in eq (1), wherein the canonical	variables provide a range of significant advantages that enhance both the analytical tractability and physical understanding of the quantum systems. This formulation of extended complex phase space is particularly useful for non-Hermitian systems where	both the potential and the effective mass function are complex, as is the case in our	study of exact solutions to the Schrödinger equation with a complex Morse potential. By extending the phase space into the complex domain, one gains access to analytic techniques such as Cauchy–Riemann conditions, which decompose the complex	Hamiltonian into two real components that correspond to a coupled real system,	allowing a deeper investigation into integrability and dynamical structure that may remain obscured in purely real formulations. The analyticity through the Cauchy–Riemann conditions on the complex Hamiltonian also reveal hidden symmetries in the two coupled real systems. Furthermore, the real and imaginary parts of the eigenfunctions and eigenspectra can be obtained explicitly, which is not the case	with any other methods applied in the literature. This aids in exploring the conditions of admitting the reality of eigenspectra for Complex potentials and investigation of their behaviour. Thus, complexifying phase space not only enriches the mathematical structure of the problem but also serves as a powerful tool for uncovering exact solvability, hidden symmetries, and physical insight in systems with complex potentials	and non-trivial mass structures. Equation \ref{eq1} can be applied  to recast the analogous  Schrödinger equation (ASE) in terms of $x_1$ and $p_2,$ 
	\begin{equation}
		\hat{H}(x,p)\psi(x) = E\psi(x).	\label{eq2}		
	\end{equation}	
	where
	\begin{equation}
		H(x,p) = -\frac{\hbar^2}{2M}\frac{\partial^2}{\partial{x^2}} + V(x). \label{eq3}
	\end{equation}
	and $V(x)$ is a complex potential. Note that since equation \ref{eq2} departs from the conventional conceptual and mathematical setting of the standard Schrodinger equation, we call equation the 'analogous Schrodinger equation' (ASE) for the non-Hermitian operator H(x,p).
	
	If the mass is considered as position dependent, the  Schrödinger equation \ref{eq2} can be written in the following form \cite{von1983position}
	\begin{equation}
		-\frac{\hbar^2}{2M}\left(\frac{\partial^2}{\partial{x}^2}-\frac{1}{M}\frac{\partial M}{\partial x}\frac{\partial}{\partial x}\right)\psi(x) = (E-V(x))\psi(x). \label{eq4}
	\end{equation}
	In the quantum context, since $p\rightarrow -i\hbar\frac{\partial}{\partial{x}}$ which implies $p_1\rightarrow -\frac{\partial}{\partial{p_2}}, x_2\rightarrow\frac{\partial}{\partial{x_1}},$ the analyticity of H(x,p) gets translated into that of the complex potential function $V(x)$.
	
	Writing the wavefunction, potential and the energy eigenvalues in real and imaginary part as 
	
	\begin{subequations} \label{eq5}
	\begin{flalign}
		& \psi(x) = \psi_r(x_1,p_2) +i\psi_i(x_1,p_2), \quad V(x) = V_r(x_1,p_2) +iV_i(x_1,p_2), \quad E = E_r + iE_i. && \label{eq5a}\\
		& \text{and considering the mass as a complex quantity with positive real part of the form} && \nonumber 
	\end{flalign}
	\begin{align}
		M &= m_r + im_i, \qquad m_r > 0, \label{eq5b} \\
		\frac{1}{M} &= \frac{1}{m_r + im_i} = \frac{m_r - im_i}{m^2}. \label{eq5c}
	\end{align}
	\end{subequations}
	
	where the subscript $r$ and $i$ respectively denote the real and imaginary parts of the corresponding physical quantity. Thus, using equations \ref{eq5a}-\ref{eq5c} and employing the analytical property of the wave function $\psi(x)$ and the mass $M$ in terms of Cauchy-Riemann condition, namely 
	
	\begin{subequations} \label{eq6}
		\begin{align}
		& \psi_{r,x_1} = \psi_{i,p_2}; \quad \psi_{r,p_2} = -\psi_{i,x_1}, \label{eq6a} \\
		& m_{r,x_1} = m_{i,p_2}; \quad m_{r,p_2} = -m_{i,x_1}. \label{eq6b}
		\end{align}
	\end{subequations}

	The partial derivative of $x$ can thus be expressed in the form of $x_1$ and $p_2$ as
	
	\begin{subequations} \label{eq7}
		\begin{align}
		\frac{\partial}{\partial{x}} &= \frac{1}{2}\left(\frac{\partial}{\partial{x_1}} - i\frac{\partial}{\partial{p_2}}\right), \label{eq7a}\\
		\frac{\partial^2}{\partial{x^2}} &= \frac{1}{4}\left(\frac{\partial^2}{\partial{x_1}^2} - \frac{\partial^2}{\partial{p_2}^2}\right) -\frac{i}{2}\frac{\partial^2}{\partial{x_1}{p_2}}. \label{eq7b}
	\end{align}
\end{subequations}

	Substituting the equations \ref{eq7a}, \ref{eq7b}, in equation \ref{eq4}
	$$-\frac{\hbar^2}{2M}\left\{\frac{1}{4}\left(\frac{\partial^2}{\partial{x_1}^2} - \frac{\partial^2}{\partial{p_2}^2}\right) -\frac{i}{2}\frac{\partial^2}{\partial{x_1}{p_2}} - \frac{1}{M}\frac{\partial M}{\partial x}\frac{\partial}{\partial x}\right\} \psi(x) = (E-V(x))\psi(x). $$
	Putting the value of mass $M$ from \ref{eq5c} in the above equation, we arrive at the following:
	
	\begin{equation} \label{eq8}
		\begin{aligned}
			& -\frac{\hbar^2}{2m^2} \bigg[ \left\{ \frac{m_r}{4} \left( \frac{\partial^2}{\partial x_1^2} - \frac{\partial^2}{\partial p_2^2} \right) - \frac{m_i}{2} \frac{\partial^2}{\partial x_1 \partial p_2} \right\}
			- i \left\{ \frac{m_i}{4} \left( \frac{\partial^2}{\partial x_1^2} - \frac{\partial^2}{\partial p_2^2} \right) + \frac{m_r}{2} \frac{\partial^2}{\partial x_1 \partial p_2} \right\} \bigg] \psi(x) \\
			&+ \frac{\hbar}{4m^4} \left[ (m_r - i m_i)^2 \frac{\partial}{\partial x_1}(m_r + i m_i) \left( \frac{\partial}{\partial x_1} - i \frac{\partial}{\partial p_2} \right) \right] \psi(x)
			= \left[ (E_r - V_r) \psi_r - (E_i \right. \\
		& \left. - V_i) \psi_i \right] + i \left[ (E_i - V_i) \psi_r + (E_r - V_r) \psi_i \right].
		\end{aligned}
	\end{equation}
	
	Solving equation \ref{eq8}, we get
	\begin{equation} \label{eq9}
		\begin{aligned}
		&\Rightarrow -\frac{\hbar^2}{2m^2}[(m_r\psi_r''+m_i\psi_i'')+i(m_r\psi_i''-m_i\psi_r'')]+\frac{\hbar^2}{2m^4}\left[\left\{(m_r^2-m_i^2)(m_r'\psi_r'-m_i'\psi_i') \right. \right. \\
		&	\left. \left. +2m_rm_i(m_i'\psi_r'+m_r'\psi_i')\right\} \right]+\frac{\hbar^2}{2m^4}\left[i\left\{(m_r^2-m_i^2)(m_i'\psi_r'+m_r'\psi_i')-2m_rm_i\left(m_r'\psi_r' \right. \right. \right. \\
		&\left. \left. \left. -m_i'\psi_i' \right) \right\} \right] = \left((E_r-V_r)\psi_r-(E_i-V_i)\psi_i\right)+ i\left((E_i-V_i)\psi_r+(E_r-V_r)\psi_i\right).
    	\end{aligned}
    \end{equation}
	The real and imaginary parts of the L.H.S. of equation \ref{eq9} can be obtained as
	
	\begin{subequations} \label{eq10}
		\begin{align}
		&	\frac{\hbar^2}{2m^4}[-m^2(m_r\psi_r''+m_i\psi_i'')+(m_r^2-m_i^2)(m_r'\psi_r'-m_i'\psi_i')+2m_rm_i(m_i'\psi_r'+m_r'\psi_i')]. \label{eq10a}\\
		&	\frac{\hbar^2}{2m^4}[-m^2(m_r\psi_i''-m_i\psi_r'')+(m_r^2-m_i^2)(m_i'\psi_r'+m_r'\psi_i')-2m_rm_i(m_r'\psi_r'-m_i'\psi_i')]. \label{eq10b}
		\end{align}
    \end{subequations}
	Defining the values of functions A,B,C,D as
	$$A = m_r\psi_r'' + m_i\psi_i'',$$
	$$B = m_r\psi_i'' - m_i\psi_r'',$$
	$$C = m_r'\psi_r' - m_i'\psi_i',$$
	$$D = m_i'\psi_r' + m_r'\psi_i'.$$
	and substituting them in equations \ref{eq10a} and \ref{eq10b}, we arrive at following equations:
	\begin{subequations} \label{eq11}
	\begin{flalign}
		& \Rightarrow \frac{\hbar^2}{2m^4}[-m^2A + (m_r^2-m_i^2)C+2m_rm_iD]= [E_r\psi_r-E_i\psi_i-V_r\psi_r+V_i\psi_i] && \label{eq11a}\\
		& \text{and} && \nonumber \\
		& \Rightarrow \frac{\hbar^2}{2m^4}[-m^2B + (m_r^2-m_i^2)D-2m_rm_iC]= [E_i\psi_r+E_r\psi_i-V_i\psi_r-V_r\psi_i]. && \label{eq11b}
	\end{flalign}
\end{subequations}
	
	Employing the eigenfunction-ansatz method, the coupled equations \ref{eq11a} and \ref{eq11b} are solved by taking the ansatz for the eigenfunction as

    \begin{subequations} \label{eq12}
	\begin{align}
		\psi(x) &= \phi(x)e^{ig(x)}. \label{eq12a} \\
		\intertext{For obtaining the solution for the ground state, we substitute}
		\phi(x) &= 1. \nonumber \\
		\intertext{Thus,}
		\psi_r + i\psi_i &= e^{i(g_r + i g_i)} = e^{-g_i}(\cos{g_r} + i\sin{g_r}). \label{eq12b} \\
		\intertext{Which implies}
		\psi_r &= e^{-g_i} \cos{g_r}; \quad \psi_i = e^{-g_i} \sin{g_r}. \label{eq12c}
	\end{align}
    \end{subequations}

	Equation \ref{eq11a} can be recast as
		\begin{equation} \label{eq13}
		\begin{aligned}
		& \frac{\hbar^2}{2m^4}\left[-m^2\left\{m_re^{-g_i}(g_i'^2\cos{g_r}-g_r'^2\cos{g_r}-g_i''\cos{g_r}-g_r''\sin{g_r}+ 2g_r'g_i'\sin{g_r})\right. \right.\\	
		& \left. +m_ie^{-g_i}(g_i'^2\sin{g_r}-g_r'^2\sin{g_r}+g_r''\cos{g_r}-g_i''\sin{g_r}-2g_r'g_i'\cos{g_r})\right\}+ (m_r^2-m_i^2)\\
		& \left\{m_r'e^{-g_i}(-g_i'\cos{g_r}-g_r'\sin{g_r})-m_i'e^{-g_i}(g_r'\cos{g_r}-g_i'\sin{g_r})\right\}+ \\
		&\left.2m_im_r\left\{m_i'e^{-g_i}(-g_i'\cos{g_r}-g_r'\sin{g_r})+m_r'e^{-g_i}(g_r'\cos{g_r}-g_i'\sin{g_r})\right\}\right] =E_re^{-g_i}\cos{g_r} \\
		&-V_re^{-g_i}\cos{g_r}-E_ie^{-g_i}\sin{g_r}+V_ie^{-g_i}\sin{g_r}. 
		\end{aligned}
       \end{equation}
	Equating the coefficient of $\cos{g_r}$ and $\sin{g_r}$ term from the equation \ref{eq13}, one obtains first set of coupled equations as
     \begin{subequations} \label{eq14}
	\begin{align}
		&\frac{\hbar^2}{2m^4}\left[-m^2\left\{m_r(g_i'^2-g_r'^2-g_i'')+m_i(g_r''-2g_r'g_i')\right\} + (m_r^2-m_i^2)\left\{-m_r'g_i'-m_i'g_r'\right\} \right. + && \nonumber\\
		&\left. 2m_rm_i\left\{m_r'g_r'-m_i'g_i'\right\} \right]=E_r-V_r, && \label{eq14a}\\
		&\frac{\hbar^2}{2m^4}\left[-m^2\left\{m_r(-g_r''+2g_r'g_i')+m_i(g_i'^2-g_r'^2-g_i'')\right\} + (m_r^2-m_i^2)\left\{-m_r'g_r'+\right. \right. && \nonumber\\
		& \left. \left. m_i'g_i'\right\}+2m_rm_i\left\{-m_i'g_r'-m_r'g_i'\right\}\right]=-E_i+V_i. && \label{eq14b}
	\end{align}
    \end{subequations}
	Similarly, substituting the values of A,B,C and D in the equation \ref{eq11b}, we get
	\begin{equation} \label{eq15}
		\begin{aligned}
		&\frac{\hbar^2}{2m^4}\left[-m^2\left\{m_re^{-g_i}(g_i'^2\sin{g_r}-g_r'^2\sin{g_r}+g_r''\cos{g_r}-g_i''\sin{g_r}-2g_r'g_i'\cos{g_r})- \right. \right. &&\\
		& \left. m_ie^{-g_i}(g_i'^2\cos{g_r}-g_r'^2\cos{g_r}-g_i''\cos{g_r}-g_r''\sin{g_r}+2g_r'g_i'\sin{g_r})\right\}+ &&\\
		&(m_r^2-m_i^2)\left\{m_i'e^{-g_i}(-g_i'\cos{g_r}-g_r'\sin{g_r})+m_r'e^{-g_i}(g_r'\cos{g_r}-g_i'\sin{g_r})\right\}- &&\\
		&\left.2m_im_r\left\{m_r'e^{-g_i}(-g_i'\cos{g_r}-g_r'\sin{g_r})-m_i'e^{-g_i}(g_r'\cos{g_r}-g_i'\sin{g_r})\right\}\right] =E_ie^{-g_i}\cos{g_r} &&\\
		&-V_ie^{-g_i}\cos{g_r}+E_re^{-g_i}\sin{g_r}-V_re^{-g_i}\sin{g_r}. &&
	\end{aligned}
\end{equation}
	Again, equating the coefficients of $\cos{g_r}$ and $\sin{g_r}$ terms respectively from equation \ref{eq15}, one obtains second set of coupled equations, namely
	\begin{subequations} \label{eq16}
		\begin{align}
		&\frac{\hbar^2}{2m^4}\left[-m^2\left\{m_r(-g_r''+2g_r'g_i')+m_i(g_i'^2-g_r'^2-g_i'')\right\} + (m_r^2-m_i^2)\left\{-m_r'g_r'+\right. \right. && \nonumber\\
		& \left. \left. m_i'g_i'\right\}+2m_rm_i\left\{-m_i'g_r'-m_r'g_i'\right\}\right]=-E_i+V_i, && \label{eq16a}\\
		&\frac{\hbar^2}{2m^4}\left[-m^2\left\{m_r(g_i'^2-g_r'^2-g_i'')+m_i(g_r''-2g_r'g_i')\right\} + (m_r^2-m_i^2)\left\{-m_r'g_i'-m_i'g_r'\right\} \right. + && \nonumber\\
		&\left. 2m_rm_i\left\{m_r'g_r'-m_i'g_i'\right\} \right]=E_r-V_r. && \label{eq16b}
		\end{align}
    \end{subequations}
	The coupled equations \ref{eq16a} and \ref{eq16b} can be solved for any complex potential by substituting the values of real and imaginary part of the given potential and taking suitable ansatz for the real and imaginary part of function $g(x)$ enumerated in equation \ref{eq12a}. In this study we wish to solve the coupled equations for complex Morse potential with PDCM.

	\section{Complex Morse potential with position-dependent complex mass} \label{sec3}
	
	In this section, the solution of ASE \ref{eq4}  for the complex Morse potential
	\begin{subequations} \label{eq17}
		\begin{align}
		 &V(x) = V_0\left[e^{-2ax}-2e^{-ax}\right]. \left(V_0, \text{a complex}\right) && \label{eq17a} 
		\intertext{is calculated in terms of parameters of extended phase space. The real and imaginary part of the potential \ref{eq17a} can be expressed as} \nonumber
		 &V_r(x_1,p_2)  = V_{0r}\left[e^{-2X}\cos{2Y}-2e^{-X}\cos{Y}\right]+V_{0i}\left[e^{-2X}\sin{2Y}-2e^{-X}\sin{Y}\right], && \nonumber \\
	     & V_i(x_1,p_2)  = V_{0i}\left[e^{-2X}\cos{2Y}-2e^{-X}\cos{Y}\right]-V_{0r}\left[e^{-2X}\sin{2Y}-2e^{-X}\sin{Y}\right]. && \label{eq17b}
	     \end{align}
    \end{subequations}
    
	where $X=a_rx_1-a_ip_2; Y= a_ix_1+a_rp_2; V_0= V_{0r}+iV_{0i}$ and $a= a_r+ia_i$ are used.
	
	The ansatz of the eigenfunction is taken in the form
	\begin{eqnarray}
		g_r(x_1,p_2) = \beta_1x_1-\alpha_1p_2+\beta_3e^{-X}\cos{Y}, \nonumber \\
		g_i(x_1,p_2) = \alpha_1x_1+\beta_1p_2-\beta_3e^{-X}\sin{Y}. \label{eq18}
	\end{eqnarray}
	
	Substituting the partial derivatives of ansatz of the eigenfunction in \ref{eq14a} and \ref{eq14b} we get
	\begin{equation} \label{eq19}
		\begin{aligned}
		& \frac{\hbar^2}{2m^4}\left[-m^2\left\{m_r\left(\alpha_1^2-\beta_1^2-e^{-2X}\cos{2Y}\beta_3^2(a_r^2-a_i^2) + e^{-X}\sin{Y}\left(2\alpha_1\beta_3a_r+2\beta_1\beta_3a_i \right. \right. \right. \right.&&\\
		& \left. \left.  +\beta_3(a_r^2-a_i^2)\right)+e^{-X}\cos{Y}(-2\alpha_1\beta_3a_i+2\beta_3\beta_1a_r-2\beta_3a_ra_i)- 2\beta_3^2a_ra_ie^{-2X}\sin{2Y} \right) &&\\
		& +m_i\left(-2\alpha_1\beta_1+e^{-X}\cos{Y}\left(2\beta_1\beta_3a_i+2\alpha_1\beta_3a_r+\beta_3(a_r^2-a_i^2)\right) e^{-X}\sin{Y}\left(2\alpha_1\beta_3a_i- \right. \right. &&  \\
		& +\left. \left.  \left. 2\beta_1\beta_3a_r+2\beta_3a_ra_i\right)+e^{-2X}\sin{2Y}\beta_3^2 (a_r^2-a_i^2)-2e^{-2X} \cos{2Y}\beta_3^2a_ra_i\right)\right\}-\left(m_r^2-m_i^2\right) &&
		\\
		& \left\{m_i'\beta_1+m_r'\alpha_1-e^{-X}\cos{Y}\left(m_i'\beta_3a_r+m_r'\beta_3a_i\right)-e^{-X}\sin{Y}\left(m_i'\beta_3a_i-m_r'\beta_3a_r\right)\right\}+ 2m_rm_i && \\
		& \left. \left\{m_r'\beta_1-m_i'\alpha_1-e^{-X}\cos{Y}\left(m_r'\beta_3a_r-m_i'\beta_3a_i\right)-e^{-X}\sin{Y}\left(m_r'\beta_3a_i+m_i'\beta_3a_r\right)\right\}\right] = E_r && \\
		& -V_{0r}\left(e^{-2X}\cos{2Y}-2e^{-X}\cos{Y}\right)-V_{0i}\left(e^{-2X}\sin{2Y}-2e^{-X}\sin{Y}\right) &&
		\end{aligned}
      \end{equation}
	and,  
		\begin{equation} \label{eq20}
		\begin{aligned}
		&\frac{\hbar^2}{2m^4}\left[-m^2\left\{m_r\left(-2\alpha_1\beta_1+e^{-X}\cos{Y}\left(2\beta_1\beta_3a_i+2\alpha_1\beta_3a_r+\beta_3(a_r^2-a_i^2)\right)+ e^{-X}\sin{Y}\right. \right. \right. &&\\
		& \left. \left(2\alpha_1\beta_3a_i-2\beta_1\beta_3a_r+2\beta_3a_ra_i\right)+e^{-2X}\sin{2Y}\left(\beta_3^2(a_r^2-a_i^2)\right)-2\beta_3^2a_ra_ie^{-2X}\cos{2Y}\right)+ m_i &&\\
		& \left(-\alpha_1^2+\beta_1^2+e^{-2X}\sin{2Y}\left(2a_r a_i \beta_3^2\right)+e^{-2X}\cos{2Y}\left(\beta_3^2(a_r^2-a_i^2)\right) +e^{-X}\sin{Y}\left(-2\alpha_1\beta_3a_r- \right. \right. &&\\
		& \left. \left. \left. 2\beta_1\beta_3a_i-\beta_3(a_r^2-a_i^2)\right)+e^{-X}\cos{Y}(2\alpha_1\beta_3a_i-2\beta_3\beta_1a_r+2\beta_3a_ra_i)-2\beta_3^2a_ra_ie^{-2X}\sin{2Y}\right)\right\}  &&\\
		& + (m_r^2-m_i^2)\left\{m_r'\beta_1-m_i'\alpha_1-e^{-X}\cos{Y}\left(m_r'\beta_3a_r-m_i'\beta_3a_i\right)- e^{-X}\sin{Y}\left(m_r'\beta_3a_i+m_i'\beta_3a_r\right)\right\}&&\\
		& +2m_rm_i \left.  \left\{m_i'\beta_1+m_r'\alpha_1-e^{-X}\cos{Y}\left(m_i'\beta_3a_r+m_r'\beta_3a_i\right)-e^{-X}\sin{Y}\left(m_i'\beta_3a_i-m_r'\beta_3a_r\right)\right\}\right] &&\\
		& =E_i-V_{0i}\left[e^{-2X}\cos{2Y}-2e^{-X}\cos{Y}\right]+V_{0r}\left[e^{-2X}\sin{2Y}-2e^{-X}\sin{Y}\right]. &&
       \end{aligned}
      \end{equation}
	Comparing both side constant terms, coefficients of $e^{-X}\cos{Y}$, $e^{-X}\sin{Y}$, $e^{-2X}\cos{2Y}$, $e^{-2X}\sin{2Y}$ of equation \ref{eq19} and \ref{eq20} respectively, we get the following non-repeating set of equations as:
	
	\begin{subequations} \label{eq21}
		\begin{align}
		&\frac{\hbar^2}{2m^4}\left[-m^2\left\{m_r\left(\alpha_1^2-\beta_1^2\right)+m_i\left(-2\alpha_1\beta_1\right)\right\}-\left(m_r^2-m_i^2\right)\left\{m_i'\beta_1+m_r'\alpha_1\right\}\right.+ 2m_rm_i && \nonumber\\
		&\left. \left\{m_r'\beta_1-m_i'\alpha_1\right\}\right] = E_r, && \label{eq21a}\\
		& \frac{\hbar^2}{2m^4}\left[-m^2\left\{m_r(-2\alpha_1\beta_1)-m_i(\alpha_1^2-\beta_1^2)\right\}+(m_r^2-m_i^2)\left\{m_r'\beta_1-m_i'\alpha_1\right\}+ 2m_rm_i\right. \nonumber &&\\
		& \left. \left\{m_i'\beta_1+m_r'\alpha_1\right\}\right] = E_i, &&\label{eq21b} \\
		&\frac{\hbar^2}{2m^4}\left[-m^2\left\{m_r\left(-2\alpha_1\beta_3a_i+2\beta_3\beta_1a_r-2\beta_3a_ra_i\right)+m_i\left(2\alpha_1\beta_3a_r+2\beta_3\beta_1a_i+\beta_3(a_r^2-a_i^2)\right) \right. \right. + &&\nonumber\\
		& \left. \left. (m_r^2-m_i^2)\left(m_i'\beta_3a_r+m_r'\beta_3a_i\right)-2m_rm_i\left(m_r'\beta_3a_r-m_i'\beta_3a_i\right)\right\}\right]=2V_{0r}, && \label{eq21c}\\
		&\frac{\hbar^2}{2m^4}\left[-m^2\left\{m_r\left(2\alpha_1\beta_3a_r+2\beta_1\beta_3a_i+\beta_3(a_r^2-a_i^2)\right)+m_i\left(2\alpha_1\beta_3a_i-2\beta_3\beta_1a_r+2\beta_3a_ra_i\right) \right. \right.+ && \nonumber\\
		& \left. \left. (m_r^2-m_i^2)\left\{-m_r'\beta_3a_r+m_i'\beta_3a_i\right\}-2m_rm_i\left(m_i'\beta_3a_r+m_r'\beta_3a_i\right)\right\}\right] = 2V_{0i}, && \label{eq21d}
	\end{align}
	\begin{align}
		&\frac{\hbar^2}{2m^4}\left[-m^2\left\{m_r\left(-2\beta_3^2a_ra_i\right)+m_i\left(\beta_3^2(a_r^2-a_i^2)\right)\right\}\right] = -V_{0i}, && \label{eq21e}\\
		&\frac{\hbar^2}{2m^4}\left[-m^2\left\{m_r\left(-\beta_3^2(a_r^2-a_i^2)\right)+m_i\left(-2\beta_3^2a_ra_i\right)\right\}\right] = -V_{0r}. && \label{eq21f}
       \end{align}
       \end{subequations}
	
	From equation \ref{eq21e} and \ref{eq21f} we obtain value of $\beta_3$ as
	\begin{subequations} \label{eq22}
	\begin{align}
		\beta_3&= \left[\frac{2m^2V_{0r}}{\hbar^2[m_r(a_r^2-a_i^2)+2m_ia_ra_i]}\right]^{1/2}, \label{eq22a}\\
		\intertext{that provides us the constraining relation among the potential parameters, namely,}	
		\frac{V_{0i}}{V_{0r}} &= \frac{2m_ra_ra_i-m_i(a_r^2-a_i^2)}{m_r(a_r^2-a_i^2)+2m_ia_ra_i}.  \label{eq22b}
       \end{align}
      \end{subequations}
	Further, equation \ref{eq21c} and \ref{eq21d} can be solved fo $\beta_1$and $\alpha_1$ to give
	\begin{subequations} \label{eq23}
	\begin{align}
		 \beta_1 &= \beta_3a_r + \frac{a_i}{2}+\frac{1}{2m^4}\left[(m_r^2-m_i^2)(m_rm_i)'+2m_rm_i(m_im_i'-m_rm_r')\right], \label{eq23a}\\
		 \alpha_1 &= \beta_3a_i - \frac{a_r}{2}+\frac{1}{2m^4}\left[(m_r^2-m_i^2)(m_im_i'-m_rm_r')-2m_rm_i(m_rm_i)'\right]. \label{eq23b}
	\end{align}
    \end{subequations}
	Defining values of $J$ and $K$ as function of masses and its derivatives,
	$$K= \left[(m_r^2-m_i^2)(m_rm_i)'+2m_rm_i(m_im_i'-m_rm_r')\right],$$
	$$J=\left[(m_r^2-m_i^2)(m_im_i'-m_rm_r')-2m_rm_i(m_rm_i)'\right].$$
	The value of $\alpha_1$ and $\beta_1$ in terms of $J$ and $K$ can be recast as
	\begin{subequations} \label{eq24}
	\begin{align}
		\beta_1 &= \beta_3a_r+\frac{a_i}{2}+\frac{K}{2m^4}, \label{eq24a}\\
		\alpha_1 &= \beta_3a_i-\frac{a_r}{2}+\frac{J}{2m^4}. \label{eq24b}
	\end{align}
\end{subequations}
	Using the results for $\beta_1$ and $\alpha_1$, we obtain the expression for the real and imaginary part of energy eigenvalue as
	\begin{subequations} \label{eq25}
	\begin{align}
		& E_r = \frac{\hbar^2}{2m^4}\left[-m^2\left\{m_r\left((\frac{1}{4}-\beta_3^2)(a_r^2-a_i^2)+\frac{1}{4m^8}(J^2-K^2)+ \beta_3(\frac{Ja_i-Ka_r}{m^4}-2a_ra_i) \right. \right. \right. &&\nonumber\\
		& \left. -\frac{1}{2m^4}(Ja_r+Ka_i)\right)-m_i\left(\frac{a_ra_i}{2}(4\beta_3^2-1)+\beta_3(a_r^2-a_i^2)+ \frac{\beta_3}{m^4}(Ja_r+Ka_i) +\frac{1}{2m^4}\left(Ja_i \right. \right. &&\nonumber\\
		& \left. \left. \left. -Ka_r \right) +\frac{JK}{2m^8}\right)\right\}-\left(m_r^2-m_i^2\right)\left\{m_i'(\beta_3a_r+\frac{a_i}{2}+\frac{K}{2m^4})+m_r'(\beta_3a_i-\frac{a_r}{2}+\frac{J}{2m^4})\right\} &&\nonumber\\
		& \left.  +2m_rm_i\left\{m_r'(\beta_3a_r+\frac{a_i}{2}+\frac{K}{2m^4})-m_i'(\beta_3a_i-\frac{a_r}{2}+\frac{J}{2m^4})\right\}\right], && \label{eq25a} \\ 
		& E_i = \frac{\hbar^2}{2m^4}\left[-m^2\left\{-m_r\left(\frac{a_ra_i}{2}(4\beta_3^2-1)+\beta_3(a_r^2-a_i^2)+\frac{\beta_3}{m^4}(Ja_r+Ka_i)+\frac{1}{2m^4} \left(Ja_i  \right.\right. \right. \right. &&\nonumber\\
		& \left. \left. -Ka_r\right)+\frac{JK}{2m^8}\right)-m_i((\frac{1}{4}-\beta_3^2)(a_r^2-a_i^2)+\frac{1}{4m^8}(J^2-K^2) +\beta_3(\frac{Ja_i-Ka_r}{m^4}-2a_ra_i) &&\nonumber\\
		& \left. -\frac{1}{2m^4}(Ja_r+Ka_i))\right\}+(m_r^2-m_i^2)\left\{m_r'(\beta_3a_r+\frac{a_i}{2}+\frac{K}{2m^4})-m_i'(\beta_3a_i-\frac{a_r}{2}+\frac{J}{2m^4})\right\} && \nonumber\\
		& \left. + 2m_r'm_i'\left\{m_i'(\beta_3a_r+\frac{a_i}{2}+\frac{K}{2m^4})+m_r'(\beta_3a_i-\frac{a_r}{2}+\frac{J}{2m^4})\right\}\right]. && \label{eq25b} 	
	\end{align}
    \end{subequations}
	And the eigenfunction is expressed into the following form as
	\begin{equation}
		\psi = \exp\left[\left\{(\frac{1}{2}+i\beta_3)a-\frac{J-iK}{2m^4}\right\}x+i\beta_3e^{-ax}\right]. \label{eq26}
	\end{equation}
	
	\section{Condition for normalized eigenfunction and probability density}\label{sec4}
	
	The normalization of the eigenfunction of the PDCM system with
	complex Morse’s potential, like other non-Hermitian quantum systems, is still an open problem
	and subject of research and discussion in the literature. The normalized eigenfunction for the
	said potential can be written from eq\ref{eq26} as
	\begin{equation}
		\psi\left(x\right) = N \exp\left[\left\{(\frac{1}{2}+i\beta_3)a-\frac{J-iK}{2m^4}\right\}x+i\beta_3e^{-ax}\right] \label{eq27}
	\end{equation} 
	Where, $N$ is the normalization constant. 
	
	The condition for the normalization of eigenfunction can be expressed as, 
	\begin{equation}
		\begin{aligned}
			&\int_{-\infty}^{\infty} \int_{-\infty}^{\infty} N^2 \psi^*\left(x_1,p_2\right) \psi\left(x_1,p_2\right) dx_1 dp_2 =N^2\int_{-\infty}^{\infty} \int_{-\infty}^{\infty} |\psi\left(x_1,p_2\right)|^2 dx_1 dp_2 = 1 &&\label{eq28}
		\end{aligned}
	\end{equation}
	where, $\psi^*\left(x_1,p_2\right)$ represents the complex conjugate of the eigenfunction. 
	
	The integral defined in equation \ref{eq28} can be reduced in terms $\alpha_1$ and $\beta_1$ as,
	\begin{equation}
		\begin{aligned}
		& N^2\int_{-\infty}^{\infty} \int_{-\infty}^{\infty} |\psi\left(x_1,p_2\right)|^2 dx_1 dp_2 = N^2\int_{-\infty}^{\infty}\int_{-\infty}^{\infty}\exp\left[-2\left\{\alpha_1 x_1 + \beta_1 p_2\right\}\right] dx_1 dp_2 = 1. &&\label{eq29}
		\end{aligned}
	\end{equation}
	The above two dimensional integral is divergent for the range of $x_1$ and $p_2$ in the region $\left(-\infty,\infty\right)$. It is interesting to observe that the above integral converges only for the absolute values of $x_1$ and $p_2$ provided both $\alpha_1$ and $\beta_1$ are considered positive values. Thus, by considering $x_1$ and $p_2$ as absolute values, and putting the value of integrals
	$$
	\int_{-\infty}^{\infty} \exp[-2\left\{\alpha_1|x_1|\right\}] dx_1 = \frac{1}{\alpha_1}, \quad 	\int_{-\infty}^{\infty} \exp[-2\left\{\beta_1|p_2|\right\}] dp_2 = \frac{1}{\beta_1};\quad \alpha_1>0, \beta_1>0 
	$$
	the two dimensional integral in equation \ref{eq29} takes the form,
	\begin{equation}
		N^2 \int_{-\infty}^{\infty} \int_{-\infty}^{\infty} |\psi\left(x_1,p_2\right)|^2 dx_1 dp_2 = N^2 \frac{1}{\alpha_1\beta_1} = 1. \label{eq30}
	\end{equation}
	Which implies that the normalization constant is,
	\begin{equation}
		N = \left(\alpha_1 \beta_1\right)^\frac{1}{2}. \label{eq31}
	\end{equation}
	The constraints on value of both $\alpha_1$ and $\beta_1$ to admit positive values can be recast in terms of $\beta_3$ as,
	
	\begin{equation}
		\beta_3 > -\frac{a_i}{2 a_r} -\frac{K}{2 m^4 a_r}, \label{eq32}
	\end{equation}
	\begin{equation}
		\beta_3 > \frac{a_r}{2 a_i} - \frac{J}{2 m^4 a_i}. \label{eq33}
	\end{equation}
	
	The normalized eigenfunction and probablity density function acquire the following form:
	\begin{eqnarray}
		\psi\left(x\right) = \sqrt{\alpha_1 \beta_1} \exp\left[\left\{(\frac{1}{2}+i\beta_3)a-\frac{J-iK}{2m^4}\right\}x+i\beta_3e^{-ax}\right], \label{eq34} \\
		\psi^*\left(x_1,p_2\right) \psi\left(x_1,p_2\right) = \alpha_1 \beta_1 \exp\left[-2\left\{\alpha_1 |x_1| + \beta_1 |p_2|\right\}\right]. \label{eq35}
	\end{eqnarray}
	
	\section{Analysis of eigenvalues and eigenfunctions for the general case of position-dependent complex mass} \label{sec5}
	
	To study the nature of eigenfunctions and eigenvalues admitted by the PDCM, we recall that PDCM is a linear and analytical function of $x$ satisfying the C-R equations for the real and imaginary part of the mass defined in eq.\ref{eq6b}. The most general case of such mass profile can be expressed in the following form: 
\begin{subequations} \label{eq36}
	\begin{align}
		m_r &= c x_1 - d p_2 + e_1, \label{eq36a} \\
		m_i &= d x_1 + c p_2 + e_2. \label{eq36b} 
	\end{align}
	\end{subequations}
	where, $c$, $d$, $e_1$ and $e_2$ are real constants. 
	
\begin{figure}[h!]
		\centering
		\fbox{\includegraphics[scale=0.4]{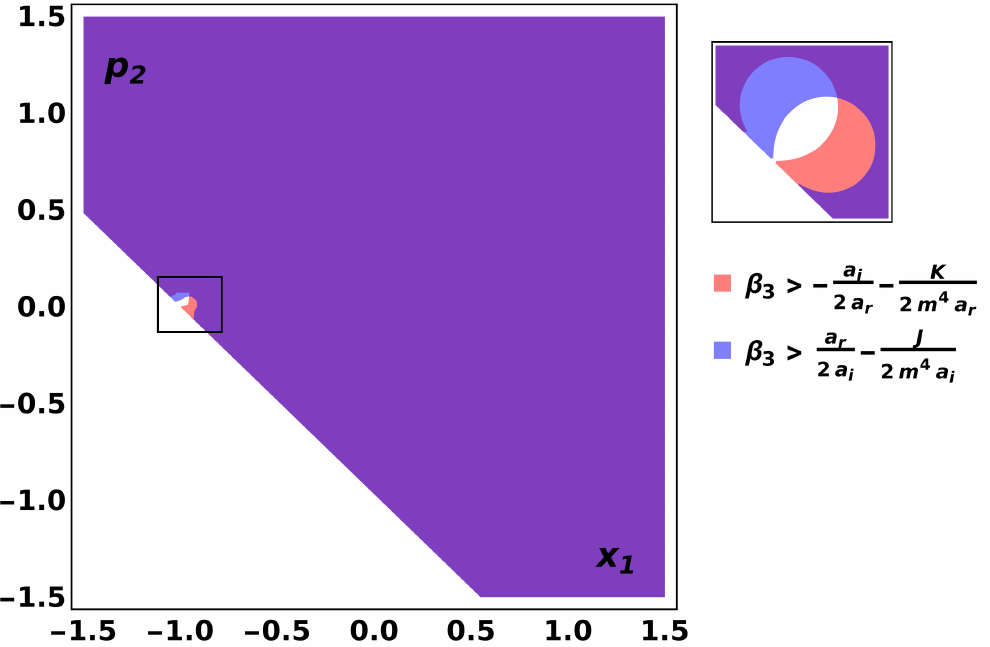}}
		\captionsetup{aboveskip=4pt}
		\caption{The Normalization condition plot for a position-dependent complex mass system.  Here, the purple shade defines the intersection region of both the normalization conditions, whereas white, blue and pink shades show the region where eigenfunction can’t be normalized.}
		\label{fig1}
	\end{figure}

	Substituting the value of the masses the expression for the ansatz parameters $\beta_1$, $\alpha_1$ and $\beta_3$ can be obtained. These parameters can then further be substituted in equations. \ref{eq32} and \ref{eq33} to derive constraints on the normalization conditions of the eigenfunction. Figure \ref{fig1} shows the plot of $\beta_3$  with respect to $x_1$ and $p_2$. The plot simply reveals the possible regions in the argand plane to achieve normalized eigenfunctions of the complex Morse potential and hence gives clues to the existence of the quantum system considered. 
	
\begin{figure*}[h!]
		\centering
		\begin{subfigure}[b]{0.4\paperwidth}
			\centering
			\fbox{\includegraphics[scale=0.36]{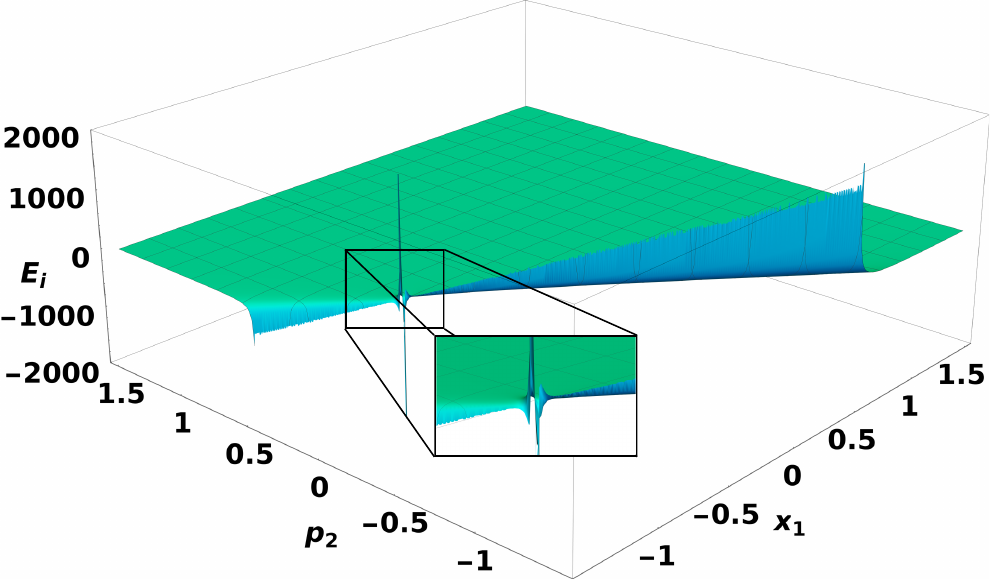}}
			\caption{}
			\label{fig2a}
		\end{subfigure}
		\hspace*{-5em}
		\begin{subfigure}[b]{0.41\paperwidth}
			\centering
			\fbox{\includegraphics[scale=0.36]{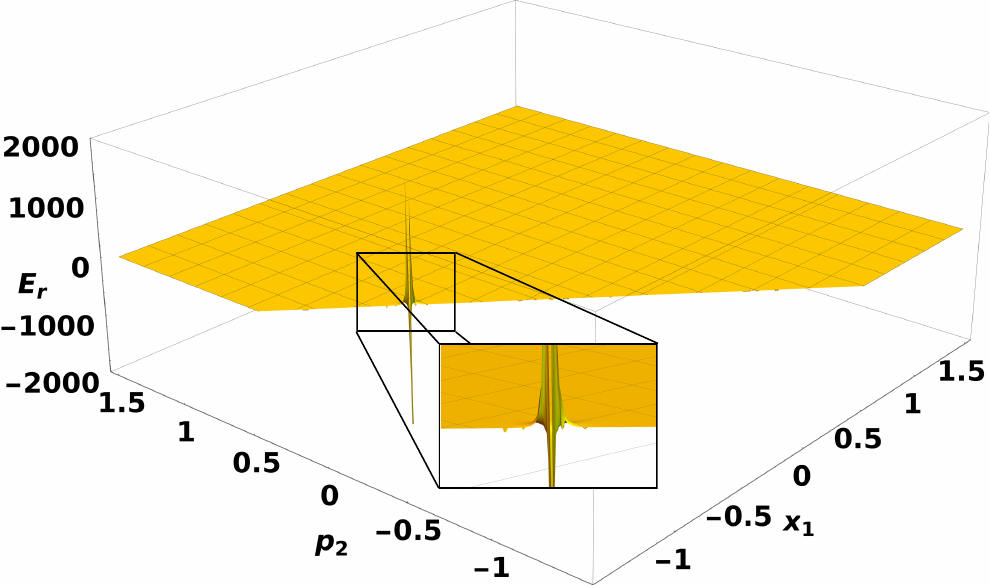}}
			\caption{}
			\label{fig2b}
		\end{subfigure}
		\captionsetup{aboveskip=4pt}
		\caption{The eigenvalue plots of mass profile $m_r = cx_1-dp_2+e_1$ and $m_i = dx_1+cp_2+e_2$ in extended complex plane. (a) $E_i$ Vs $x_1$ and $p_2$ plot (b) $E_r$ Vs $x_1$ and $p_2$.}
		\label{fig2} 	 
	\end{figure*}

	On substituting the values of the real and imaginary part of the PDM defined in equations \ref{eq36a} and \ref{eq36b} into equations \ref{eq25a} and \ref{eq25b}, the real and imaginary part of the eigenvalues can be evaluated which turn out to be quite lengthy. The nature and behaviour of these eigenvalues is studied by plotting them in the Argand plane with respect to $x_1$ and $p_2$ (figures \ref{fig2a} and \ref{fig2b}). 
	
	\begin{figure*}[h!]
		\centering
		\begin{subfigure}[b]{0.4\paperwidth}
			\centering
			\fbox{\includegraphics[scale=0.4]{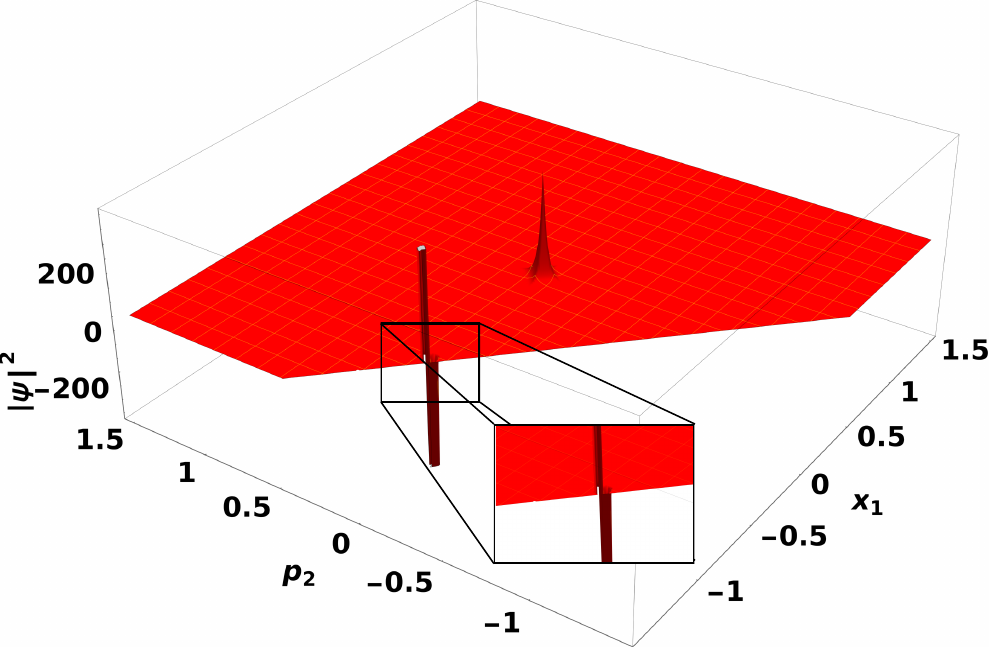}}
			\caption{}
			\label{fig3a}
		\end{subfigure}
		\hspace*{-5em}
		\begin{subfigure}[b]{0.4\paperwidth}
			\centering
			\fbox{\includegraphics[scale=0.47]{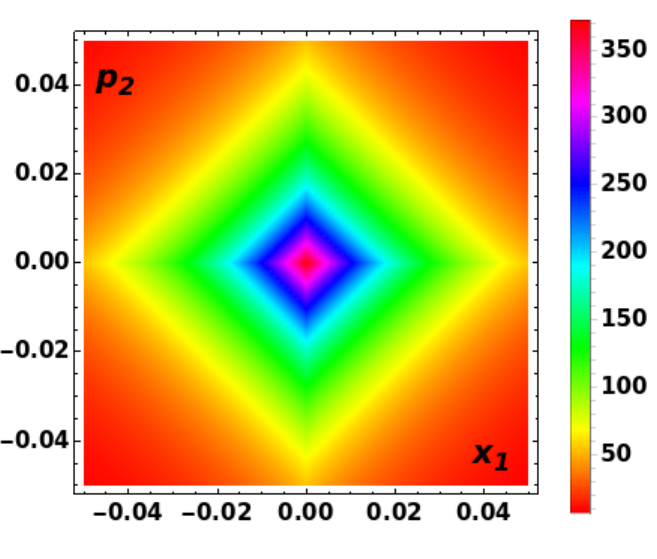}}
			\caption{}
			\label{fig3b}
		\end{subfigure}
		
		\bigskip
		
		\begin{subfigure}[b]{0.4\paperwidth}
			\centering
			\fbox{\includegraphics[scale=0.3]{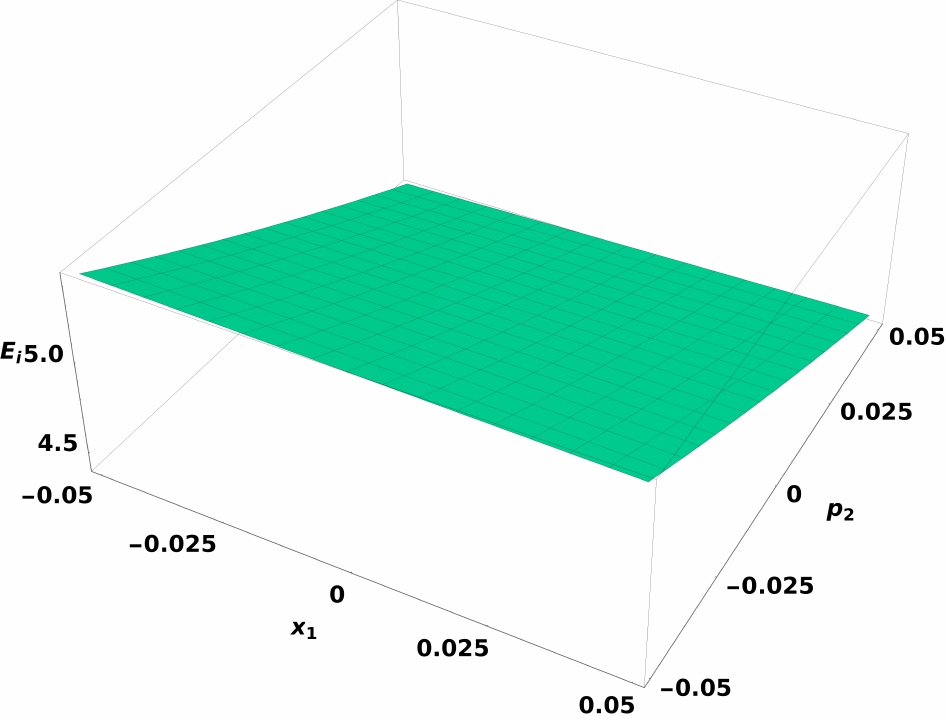}}
			\caption{}
			\label{fig3c}
		\end{subfigure}
		\hspace*{-5em}
		\begin{subfigure}[b]{0.4\paperwidth}
			\centering
			\fbox{\includegraphics[scale=0.32]{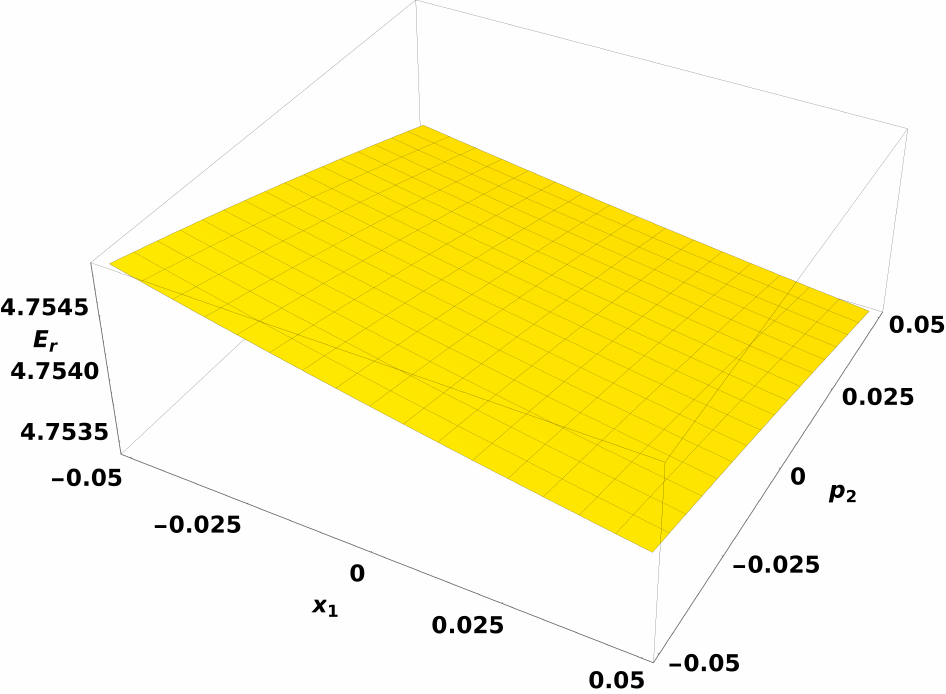}}
			\caption{}
			\label{fig3d}
		\end{subfigure}
		\captionsetup{aboveskip=4pt}
		\caption{The general case plots for mass profile $m_r = cx_1-dp_2+e_1$ and $m_i = dx_1+cp_2+e_2$ in extended complex plane. (a) Probability density (PCD) Plot (b) Contour plot of peak in figure \ref{fig3a} (c) Behaviour of Imaginary part of eigenvalue at peak of PCD (d) Behaviour of real part of eigenvalue at peak of PCD. }
		\label{fig3} 	 
	\end{figure*} 
	
	The probability density can be evaluated from equation \ref{eq35} and is plotted in figure \ref{fig3a} and \ref{fig3b}. Interestingly, the real and imaginary values of the eigenvalues of the Complex Morse potential plotted in figures \ref{fig2} and \ref{fig3} subscribe to the characteristics of the eigenfunction of the said potential enumerated in figure \ref{fig1}, where the eigenvalues are only defined in the region where the normalization conditions of the eigenfunctions are satisfied. 
	
	Figures \ref{fig2a} and \ref{fig2b} illustrate the variation of the imaginary and real part of the eigenvalue in the Argand plane as a function of $x_1$ and $p_2$. As expected, for a PDCM system, the eigenvalues vary with position. In the region where only one normalization conditions is fulfilled, highlighted in pink and blue region in figure \ref{fig1}, the eigenvalue plots exhibit anomalous behavior, with the ground state energy diverging to either infinitely large positive or negative values. Figures \ref{fig3c} and \ref{fig3d} display the positive ground state eigenvalue in the region corresponding to the peak observed in the probability current density plot (figure \ref{fig3a}). In figure \ref{fig3a} this region, where only a single normalization condition holds, is specifically marked, and it is noteworthy that the probability density in this area diverges to infinitely positive or negative values. The same plot also indicates the domain in which the diatomic complex system is likely to exist, as evidenced by the peak in probability current density. To further analyze this behavior,  figure \ref{fig3b} provides a contour plot showing an increased probability of locating the particle near the origin.

	\subsection{Special Cases of Mass Profile}
	To achieve a better insight into the behaviour of the complex Morse potential with PDCM, let us analyze the special cases where masses are constant along the real or imaginary axis. 
	
	\subsubsection{Case I(a), when $\frac{dm_r}{dx_1} = 0$} \label{case Ia}
	
	\hfil
	
	For the case where the real part of the PDCM does not vary along the real axis, the value of $\alpha_1$ , $\beta_1$ and the general expression of real and imaginary eigenvalues are obtained as, 

		\begin{align}
		&\alpha_1 = a_i \beta_3 - \frac{a_r}{2} - \frac{m_i m'_i}{2m^2}, &&\label{eq37} \\
		&\beta_1 = a_r \beta_3 + \frac{a_i}{2} + \frac{m_r m'_i}{2m^2}, &&\label{eq38} \\
		& E_i = \frac{1}{8 m^6}\left[m^4\left\{(a_r^2 - a_i^2)((1-4\beta_3^2) m_i - 4 \beta_3 m_r) - 2 a_i a_r((1-4 \beta_3^2)m_r + 4 \beta_3 m_i)\right\} + \right. &&\nonumber\\
		& \left.  (3m_r^2 m_i - m_i^3){m'_i}^2\right], \label{eq39} &&\\
		& E_r = \frac{1}{8 m^6}\left[m^4\left\{(a_r^2 - a_i^2)((4\beta_3^2-1) m_r - 4 \beta_3 m_i) + 2 a_i a_r((4 \beta_3^2 -1)m_i + 4 \beta_3 m_r)\right\} + \right. &&\nonumber\\
		& \left.  (3m_r m_i^2 - m_r^3){m'_i}^2\right]. &&\label{eq40}
		\end{align}
	
	Substituting the values of $\alpha_1$ and $\beta_1$ from equations \ref{eq37} and \ref{eq38} in equations \ref{eq32}, \ref{eq33} and \ref{eq35}, the condition of normalization and probability density can be obtained. The mass function defined in equation \ref{eq36a} and \ref{eq36b} for this case is expressed in the form, 
	\begin{subequations} \label{eq41}
	\begin{align}
		m_r &= -d p_2 + e_1, \label{eq41a} \\
		m_i &= d x_1 + e_2. \label{eq41b}
	\end{align}
\end{subequations}
	
	\subsubsection{Case II(a), when $\frac{dm_i}{dx_1} = 0$} \label{case IIa}
	
	\hfil
	
	Let us consider the case when the imaginary part of the mass is independent of $x_1$. The values of the parameters $\alpha_1$ and $\beta_1$ along with the real and imaginary parts of the eigenvalues are obtained in the following form:
	\begin{align}
		&\alpha_1 = a_i \beta_3 - \frac{a_r}{2} - \frac{m_r m'_r}{2m^2}, &&\label{eq42} \\
		&\beta_1 = a_r \beta_3 + \frac{a_i}{2} - \frac{m_i m'_r}{2m^2}, &&\label{eq43} \\
		& E_i = \frac{1}{8 m^6}\left[m^4\left\{(a_r^2 - a_i^2)((1-4\beta_3^2) m_i - 4 \beta_3 m_r) - 2 a_i a_r((1-4 \beta_3^2)m_r + 4 \beta_3 m_i)\right\} - \right. && \nonumber\\
		& \left.  (3m_r^2 m_i - m_i^3){m'_r}^2\right], && \label{eq44} \\
		& E_r = \frac{1}{8 m^6}\left[m^4\left\{(a-r^2 - a_i^2)((4\beta_3^2-1) m_r - 4 \beta_3 m_i) + 2 a_i a_r((4 \beta_3^2 -1)m_i + 4 \beta_3 m_r)\right\} - \right. &&\nonumber\\
		& \left.  (3m_r m_i^2 - m_r^3){m'_r}^2\right]. &&\label{eq45}
	\end{align}
	The mass function for this case is written as,
	\begin{subequations} \label{eq46}
		\begin{align}
		m_r = c x_1 + e_1, \label{eq46a} \\
		m_i = c p_2 + e_2. \label{eq46b}
		\end{align}
\end{subequations}

	\subsubsection{Comparisons of Cases I(a) and II(a)}
	
	\hfil
	
	\begin{enumerate}[label=(\roman*)]
		\item Normalization condition
		
		Let us compare the behaviour of particles with PDCM under the action of complex Morse potential. The expressions for the normalization condition of the eigenfunction admitted in the above mentioned special cases defined in \ref{case Ia} and \ref{case IIa} are derived and plotted in figures \ref{fig4a} and \ref{fig4b}.

		\begin{figure*}[h!]
			\centering
			\begin{subfigure}[b]{0.4\paperwidth}
				\centering
				\fbox{\includegraphics[scale=0.38]{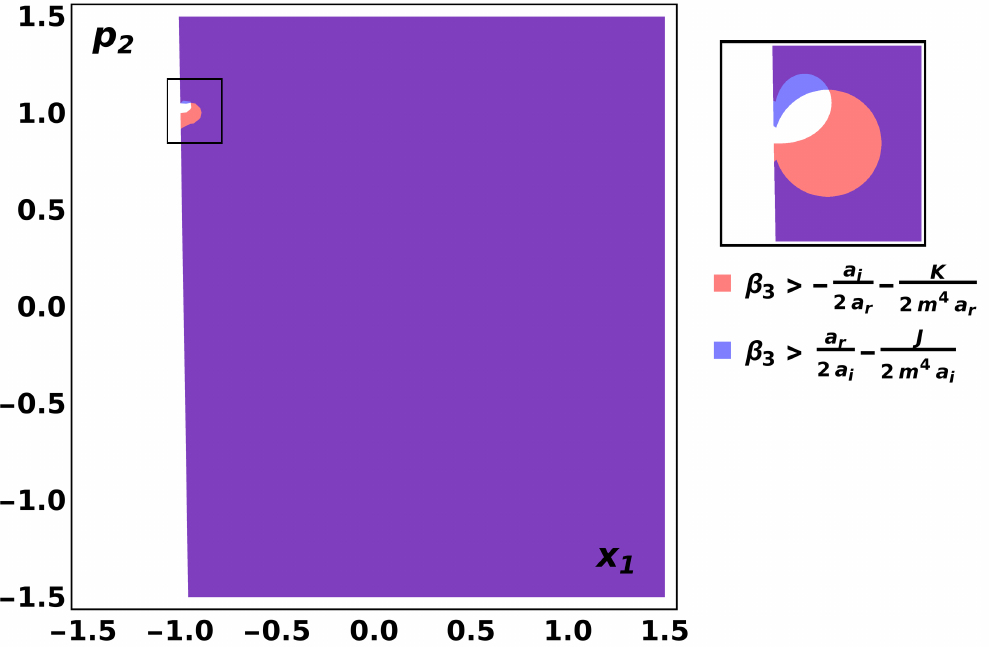}}
				\caption{}
				\label{fig4a}
			\end{subfigure}
			\hspace*{-5em}
			\begin{subfigure}[b]{0.4\paperwidth}
				\centering
				\fbox{\includegraphics[scale=0.38]{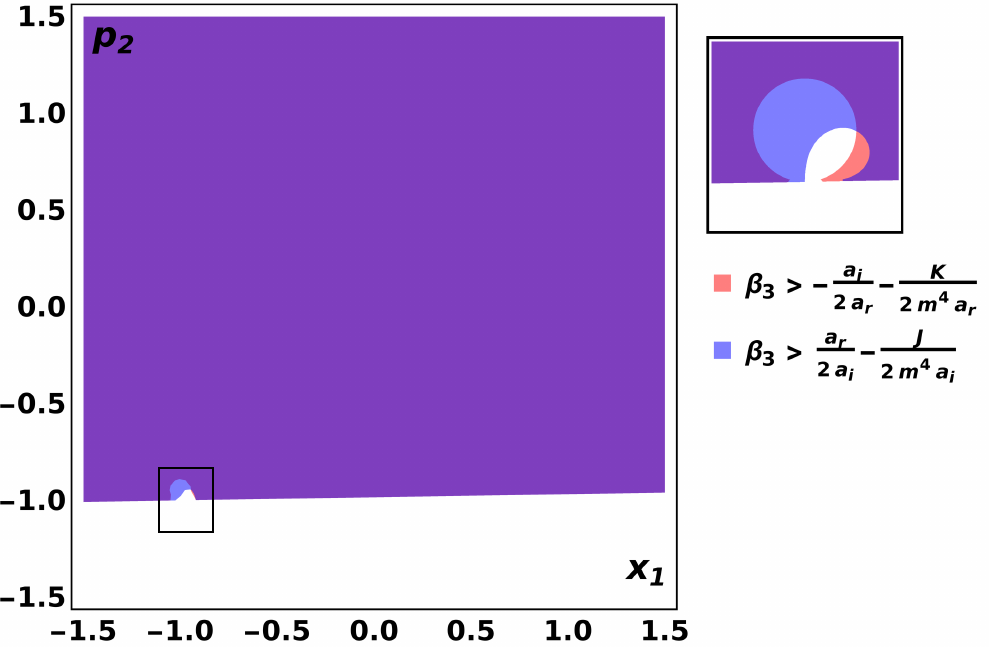}}
				\caption{}
				\label{fig4b}
			\end{subfigure}
			\captionsetup{aboveskip=4pt}
			\caption{Normalization condition plot for (a) Case I(a) (b) Case II(a)}
			\label{fig4} 	 
		\end{figure*} 
		
		The comparison of the general case enumerated in figure \ref{fig1} with the special cases plotted in figure \ref{fig4} highlights the different regions of the Argand plane where the eigenfunction is normalized.
		
		\item Behaviour of Eigenvalues
		
		The blue and pink regions highlighted in figures \ref{fig4a} and \ref{fig4b}, which satisfy only one of the two required conditions for the existence of normalized eigenfunctions, display divergent behavior in the real and imaginary parts of the eigenvalues. This is evident from the zoomed-in views in figures \ref{fig5a}-\ref{fig5d}. Therefore, the system’s behavior in these regions aligns with the general case of position-dependent complex mass described by Equations \ref{eq36a} and \ref{eq36b}.
		
		\begin{figure*}[h!]
			\centering
			\begin{subfigure}[b]{0.4\paperwidth}
				\centering
				\fbox{\includegraphics[scale=0.4]{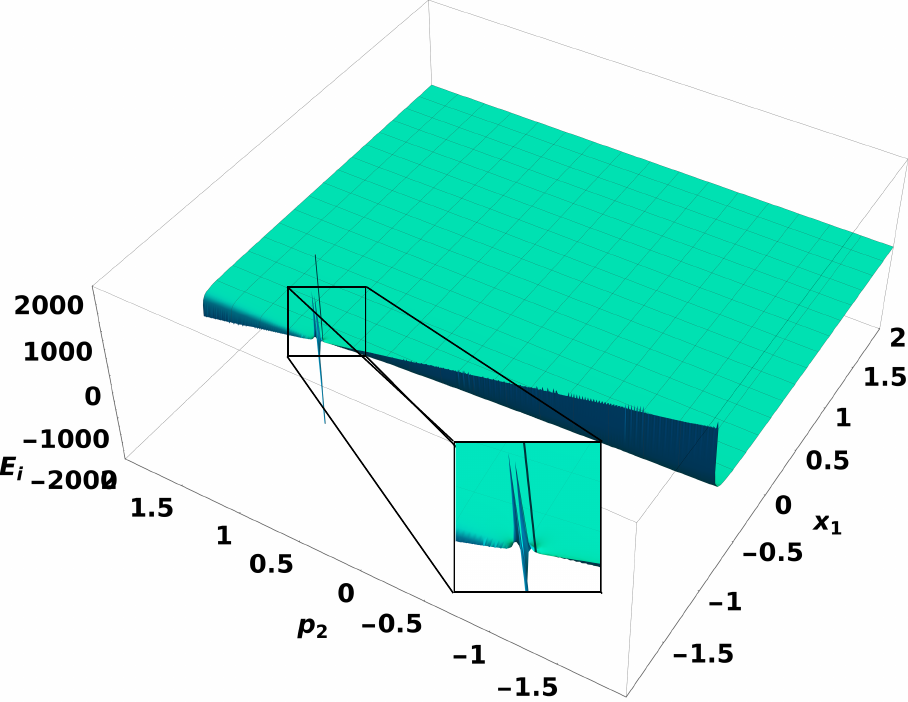}}
				\caption{}
				\label{fig5a}
			\end{subfigure}
			\hspace*{-5em}
			\begin{subfigure}[b]{0.4\paperwidth}
				\centering
				\fbox{\includegraphics[scale=0.4]{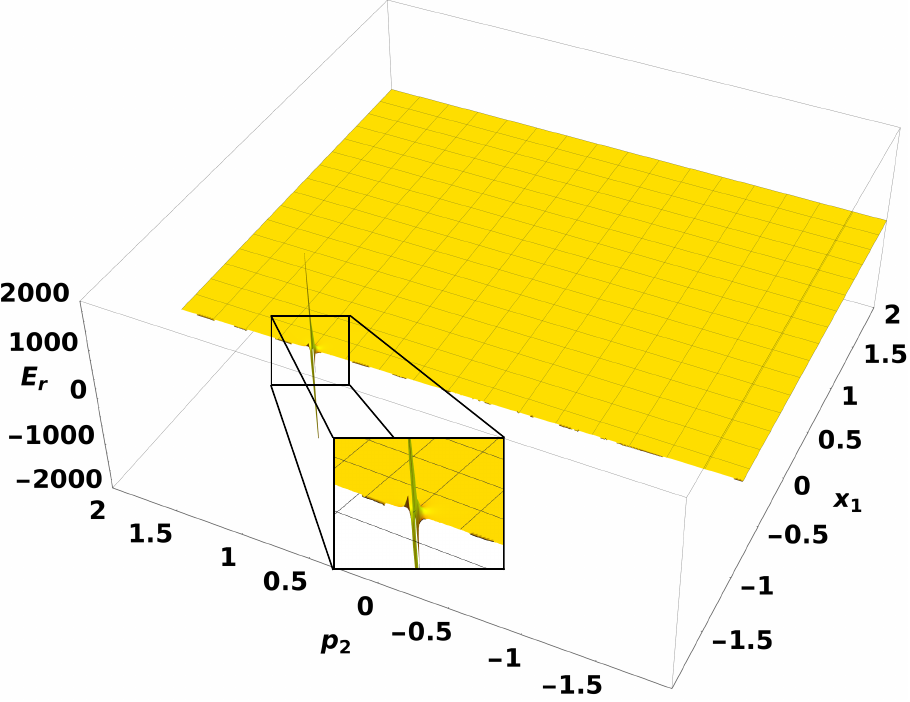}}
				\caption{}
				\label{fig5b}
			\end{subfigure}
			
			\bigskip
			
			\begin{subfigure}[b]{0.4\paperwidth}
				\centering
				\fbox{\includegraphics[scale=0.38]{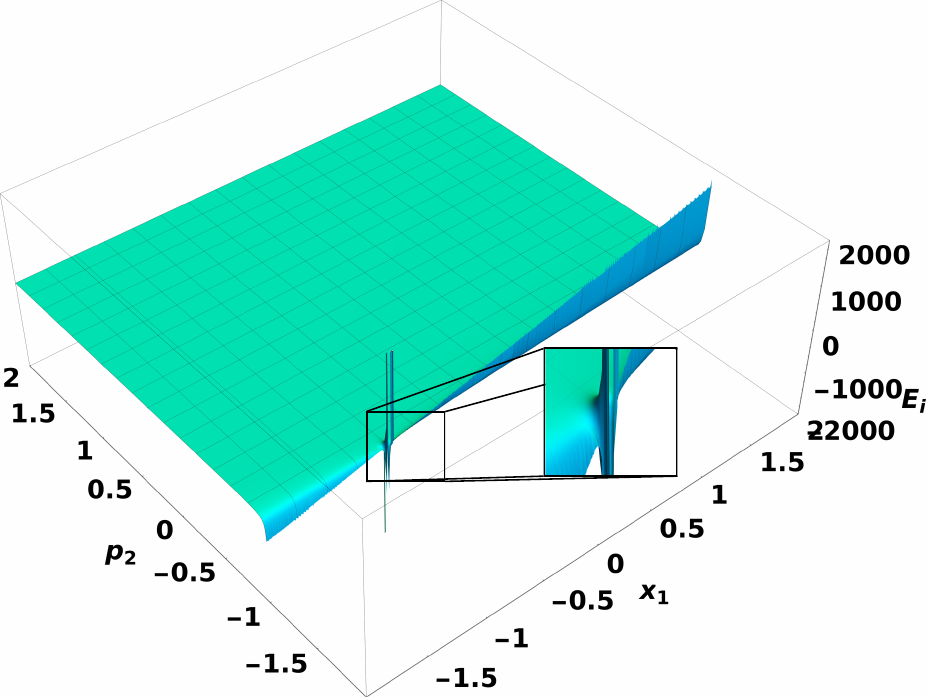}}
				\caption{}
				\label{fig5c}
			\end{subfigure}
			\hspace*{-5em}
			\begin{subfigure}[b]{0.4\paperwidth}
				\centering
				\fbox{\includegraphics[scale=0.4]{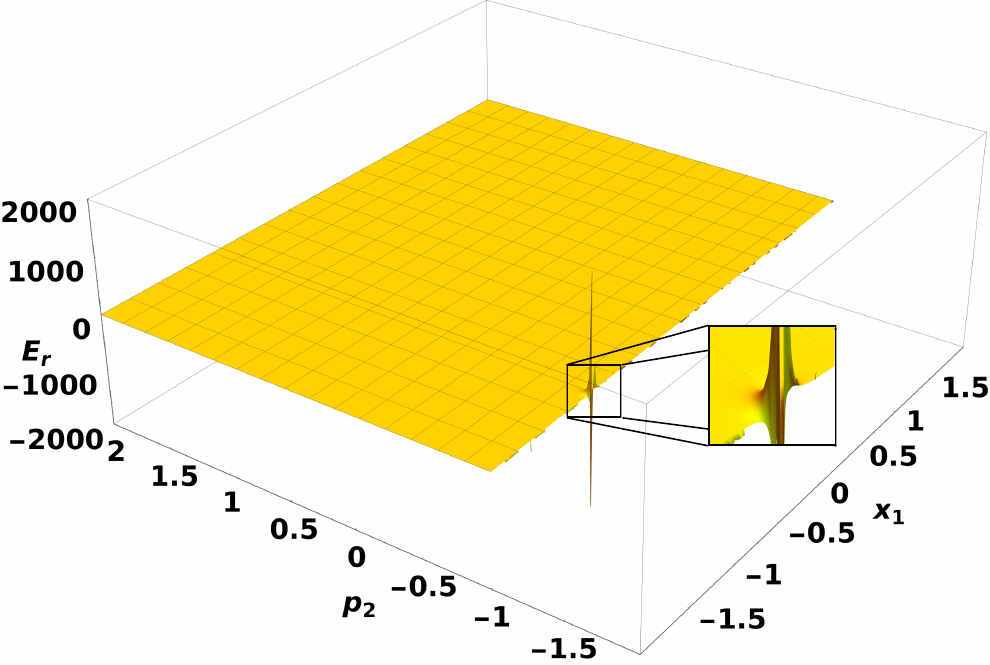}}
				\caption{}
				\label{fig5d}
			\end{subfigure}
			\captionsetup{aboveskip=4pt}
			\caption{The eigenvalue plot for Case I(a), (a) $E_i$ Vs $x_1$ and $p_2$ (b) $E_r$ Vs $x_1$ and $p_2$ and Case II(a), (c) $E_i$ Vs $x_1$ and $p_2$ (d) $E_r$ Vs $x_1$ and $p_2$.}
			\label{fig5} 	 
		\end{figure*} 
		
		\begin{figure*}[h!]
			\centering
			\begin{subfigure}[b]{0.4\paperwidth}
				\centering
				\fbox{\includegraphics[scale=0.3]{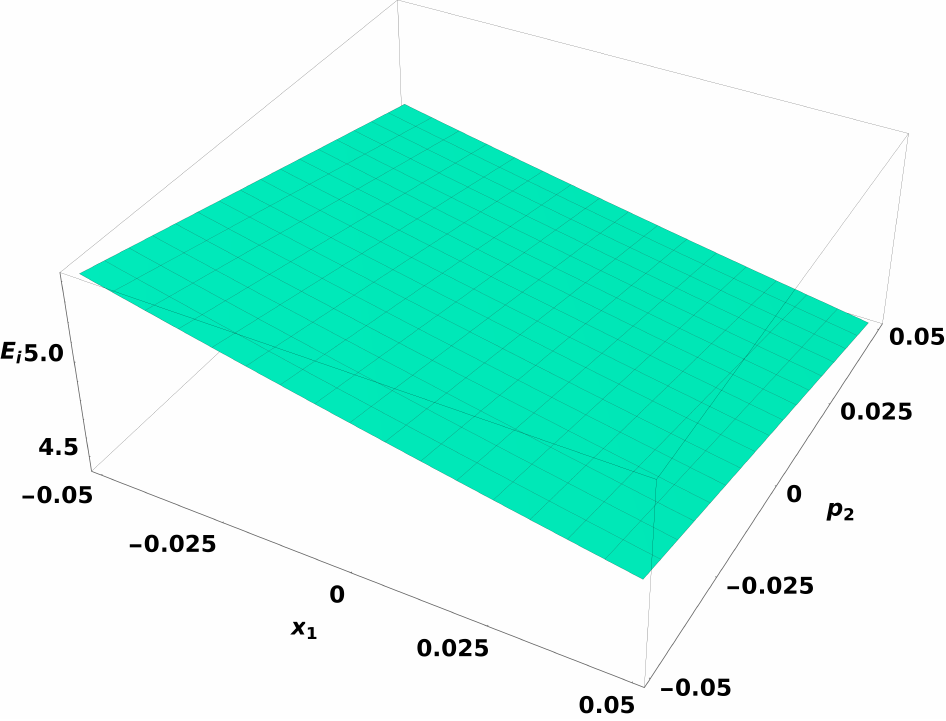}}
				\caption{}
				\label{fig6a}
			\end{subfigure}
			\hspace*{-5em}
			\begin{subfigure}[b]{0.4\paperwidth}
				\centering
				\fbox{\includegraphics[scale=0.309]{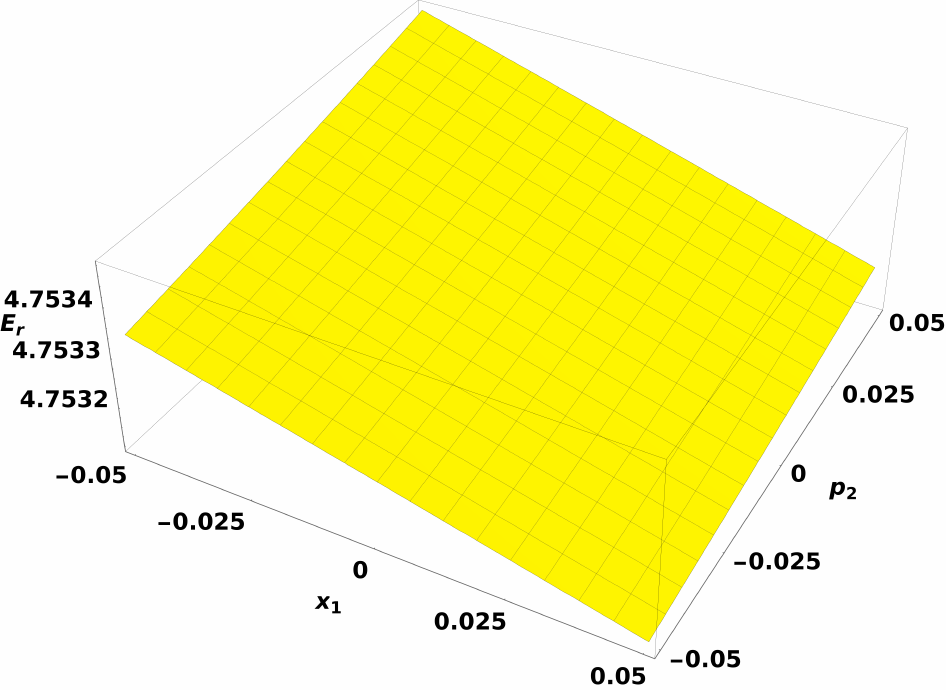}}
				\caption{}
				\label{fig6b}
			\end{subfigure}
			
			\bigskip
			
			\begin{subfigure}[b]{0.4\paperwidth}
				\centering
				\fbox{\includegraphics[scale=0.3]{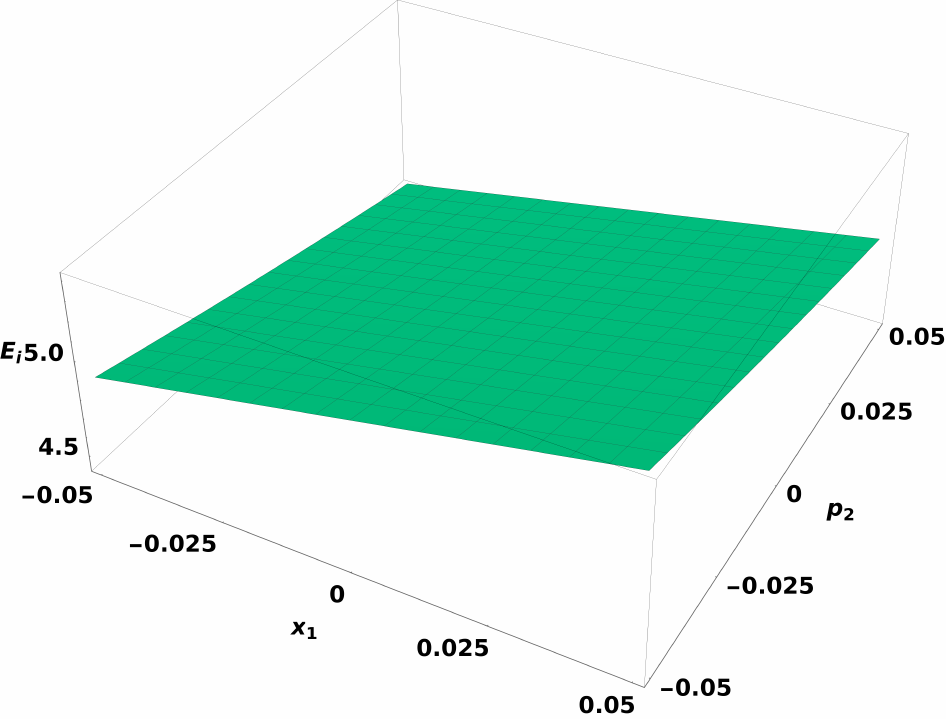}}
				\caption{}
				\label{fig6c}
			\end{subfigure}
			\hspace*{-5em}
			\begin{subfigure}[b]{0.4\paperwidth}
				\centering
				\fbox{\includegraphics[scale=0.32]{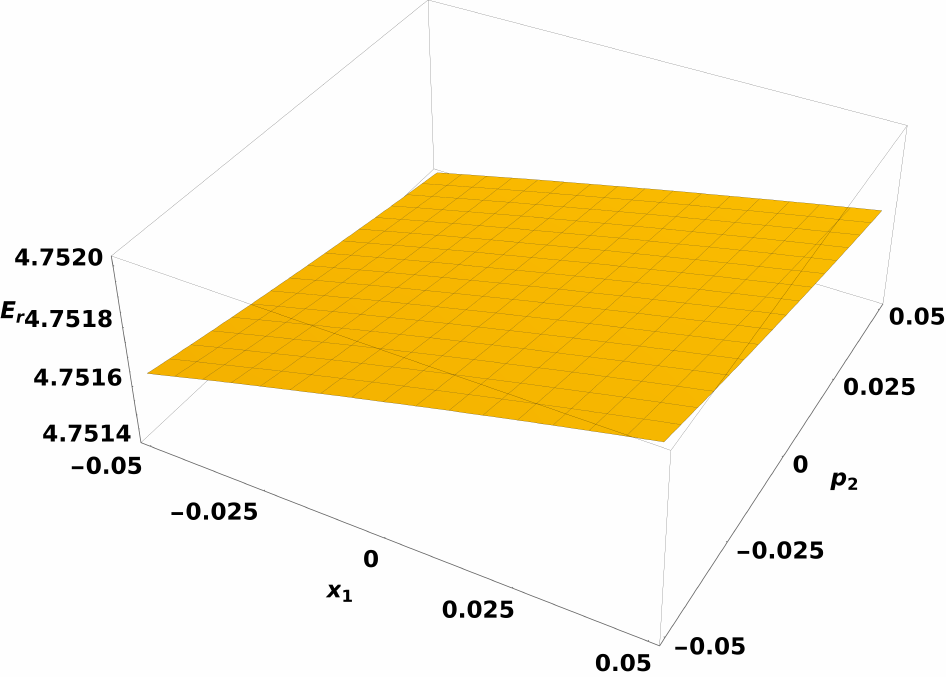}}
				\caption{}
				\label{fig6d}
			\end{subfigure}
			\captionsetup{aboveskip=4pt}
			\caption{The eigenvalue plot of the peak for Case I(a), (a) $E_i$ Vs $x_1$ and $p_2$ (b) $E_r$ Vs $x_1$ and $p_2$ and Case II(a), (c) $E_i$ Vs $x_1$ and $p_2$ (d) $E_r$ Vs $x_1$ and $p_2$.}
			\label{fig6} 	 
		\end{figure*} 
		
		\item Probability density plots
		
		The probability density plot shown in figures \ref{fig7a} and \ref{fig7b} indicates the formation of finite peaks with similar nature as in the general case. The contour plots of the region at the vicinity of the peak are demonstrated in figures \ref{fig7c} and \ref{fig7d}. The red region in the said contour plot signifies the region where the probability density is maximum and the same value decreases as one moves away from the origin. Both cases show a distinct resemblance with that of the general case (figure \ref{fig3b}).
		
		\item Behaviour of eigenvalues at peak of probability density
		
		Figure \ref{fig6} illustrates the behavior of both the real and imaginary components of the eigenvalues in the region corresponding to the peak of the probability density function for the complex diatomic molecular system, as shown in figures \ref{fig7a} and \ref{fig7b}. It is observed that the eigenvalues are confined to a limited range of positive values.
		
	\end{enumerate}
	
	\begin{figure*}[h!]
		\centering
		\begin{subfigure}[b]{0.4\paperwidth}
			\centering
			\fbox{\includegraphics[scale=0.35]{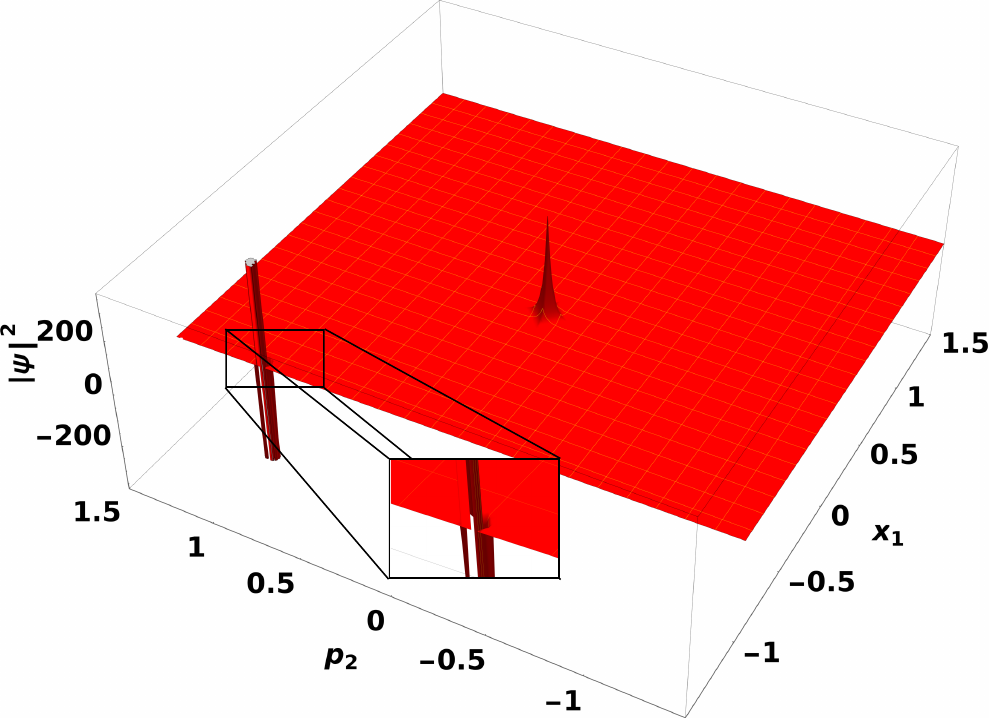}}
			\caption{}
			\label{fig7a}
		\end{subfigure}
		\hspace*{-5em}
		\begin{subfigure}[b]{0.4\paperwidth}
			\centering
			\fbox{\includegraphics[scale=0.357]{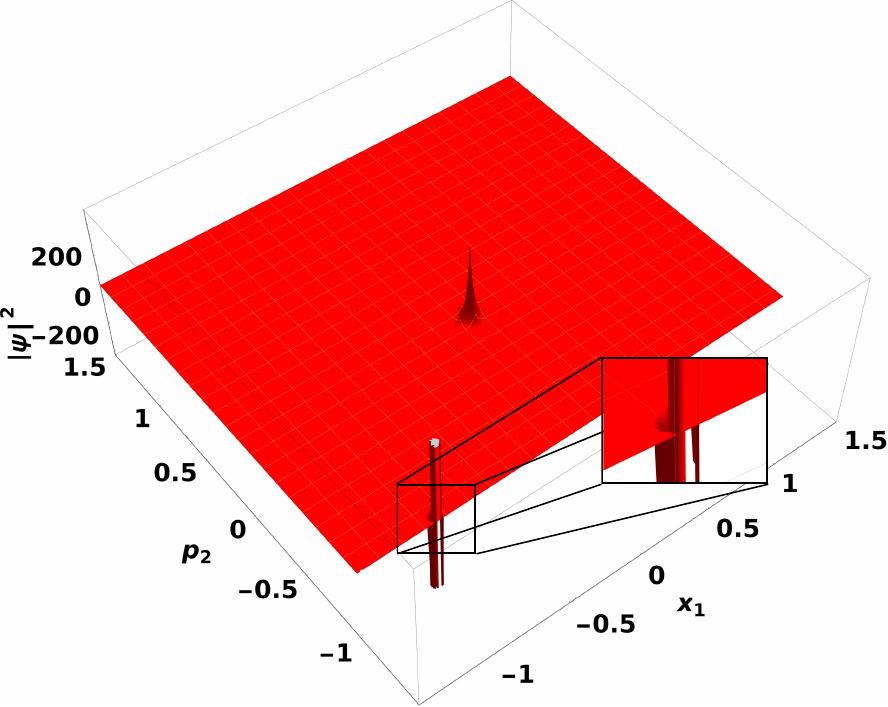}}
			\caption{}
			\label{fig7b}
		\end{subfigure}
		
		\bigskip
		
		\begin{subfigure}[b]{0.4\paperwidth}
			\centering
			\fbox{\includegraphics[scale=0.4]{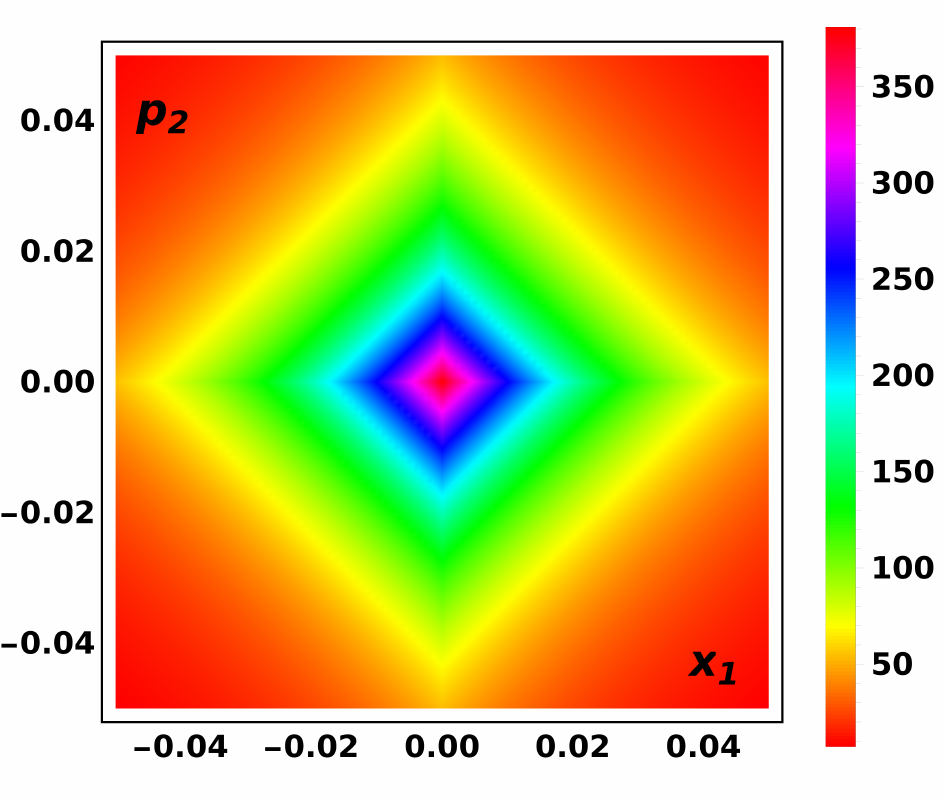}}
			\caption{}
			\label{fig7c}
		\end{subfigure}
		\hspace*{-5em}
		\begin{subfigure}[b]{0.4\paperwidth}
			\centering
			\fbox{\includegraphics[scale=0.4]{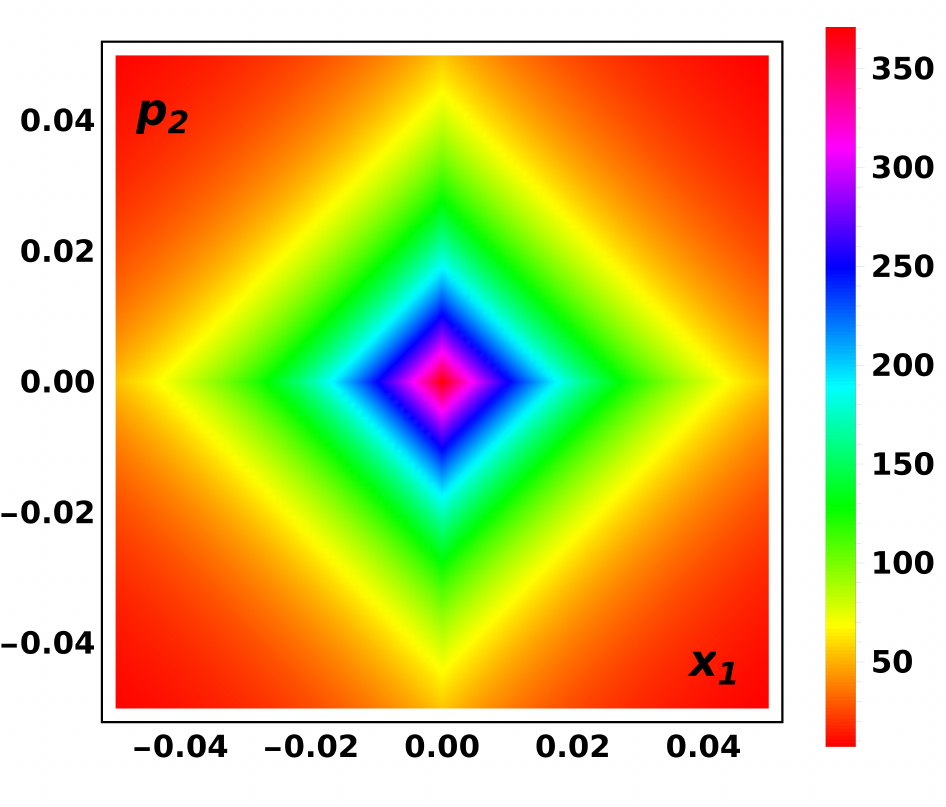}}
			\caption{}
			\label{fig7d}
		\end{subfigure}
		\captionsetup{aboveskip=4pt}
		\caption{The probablity density plots for, (a) Case I(a) (b) Case II(a) and contour plots of probability density of the peak for (c) Case I(a) (d) Case II(a)}
		\label{fig7} 	 
	\end{figure*} 
	
	\section{Reality of eigenspectrum}\label{sec6}
	
	The reality of the eigenvalue spectrum admitted by the Complex potentials has been a subject of research for the past several decades in the context of extending quantum theory beyond Hermitian operators while keeping physical observables real. The main advantage of solving the Schrödinger equation using the method described in the present study lies in the fact that the imaginary part of all physical quantities including the eigenvalues and the eigenfunction can be deduced explicitly. Equation \ref{eq12a} mandates that the energy eigenvalue admitted by the complex Morse potential is real for the ground state when
	
	\begin{equation}	
		E_i(x_1,p_2) = 0. \label{eq47}
	\end{equation}
	
	Defining $\mu$ as $\mu = \frac{m_r}{m_i}$ and writing its first-order differential with respect to $x_1$,
	$$\mu' = \frac{1}{m_i}(m'_r-\mu m'_i).$$
	The condition for the reality of the spectrum can be expressed as
	\begin{align}
		& \frac{(1+\mu)m'_i}{m_i}\left[(\alpha_1^2-\beta_1^2)+2\mu \alpha_1 \beta_1\right] \left[\frac{\mu' m_i}{m'_i} + \mu\right] \left[(\mu^2 -1)\beta_1+2\mu \alpha_1\right](\mu^2 - 1)\alpha_1+ 2 \mu \beta_1 = 0. &&\label{eq48}
	\end{align}
	
\begin{figure*}[h!]
		\centering
		\begin{subfigure}[b]{0.4\paperwidth}
			\centering
			\fbox{\includegraphics[scale=0.32]{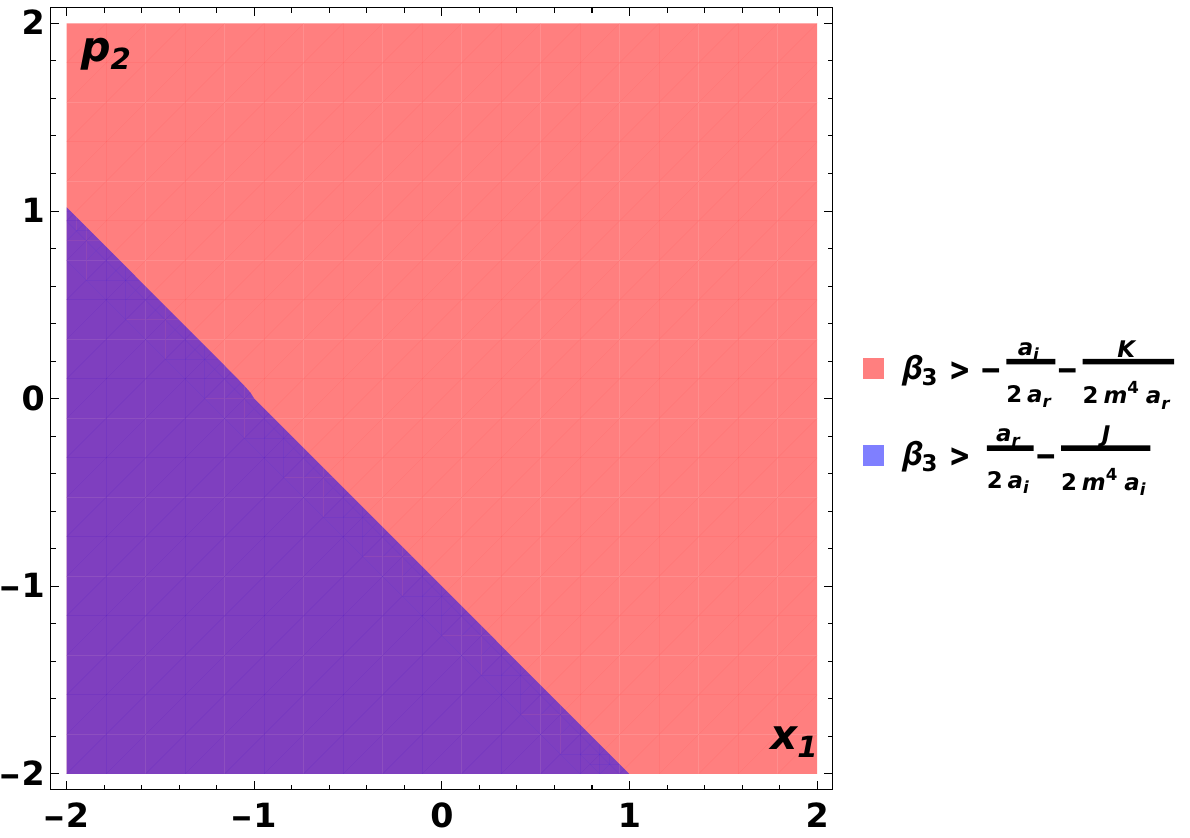}}
			\caption{}
			\label{fig8a}
		\end{subfigure}
		\hspace*{-5em}
		\begin{subfigure}[b]{0.4\paperwidth}
			\centering
			\fbox{\includegraphics[scale=0.32]{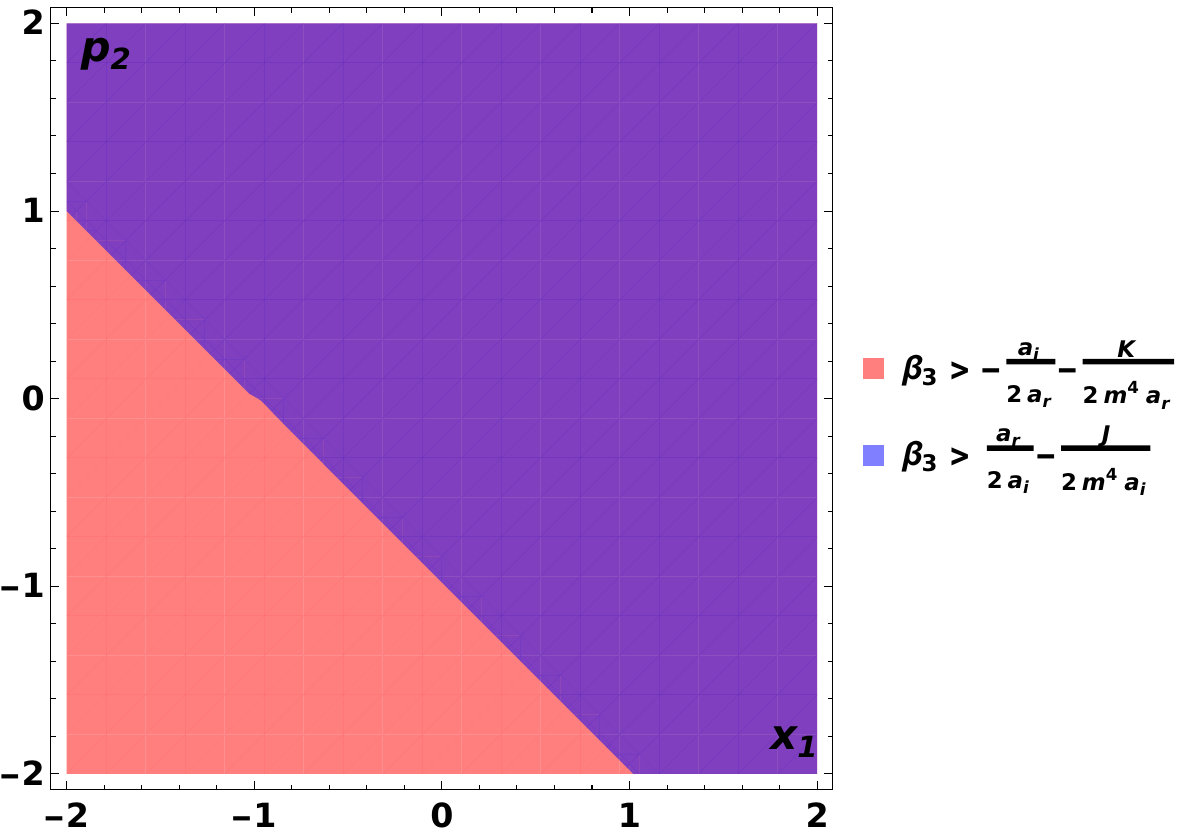}}
			\caption{}
			\label{fig8b}
		\end{subfigure}
		\captionsetup{aboveskip=4pt}
		\caption{Normalization condition plot of reality of spectrum for mass profile $m_r = cx_1-dp_2+e_1$ and $m_i = dx_1+cp_2+e_2$ in extended complex plane associated with (a) first root of $\beta_3$ (b) second root of $\beta_3$.}
		\label{fig8} 	 
	\end{figure*}

	By solving the constraint equation \ref{eq48} using Equations \ref{eq24a} and \ref{eq24b}, the values of $\alpha_1$ and $\beta_1$ are determined, which subsequently yield two distinct roots of $\beta_3$, referred to hereafter as the first and second roots of $\beta_3$. These roots are essential for ensuring the reality of the eigenvalue spectrum associated with the complex Morse potential. Additionally, the normalization conditions given in Equation \ref{eq32} and \ref{eq33} imposes further restrictions on the admissible values of $\beta_3$, ensuring that the resulting eigenvalues are real and that the corresponding probability densities are both finite and positive. Due to their complexity, the explicit forms of these roots are too cumbersome to include in the text. Figure \ref{fig8a} and \ref{fig8b} illustrate the normalization condition plots corresponding to the two distinct roots of $\beta_3$ in the position-dependent complex mass (PDCM) system. 
	
\begin{figure}[h!]
		\centering
		\begin{minipage}[b]{0.47\textwidth}
			\centering
			\fbox{\includegraphics[scale=0.4]{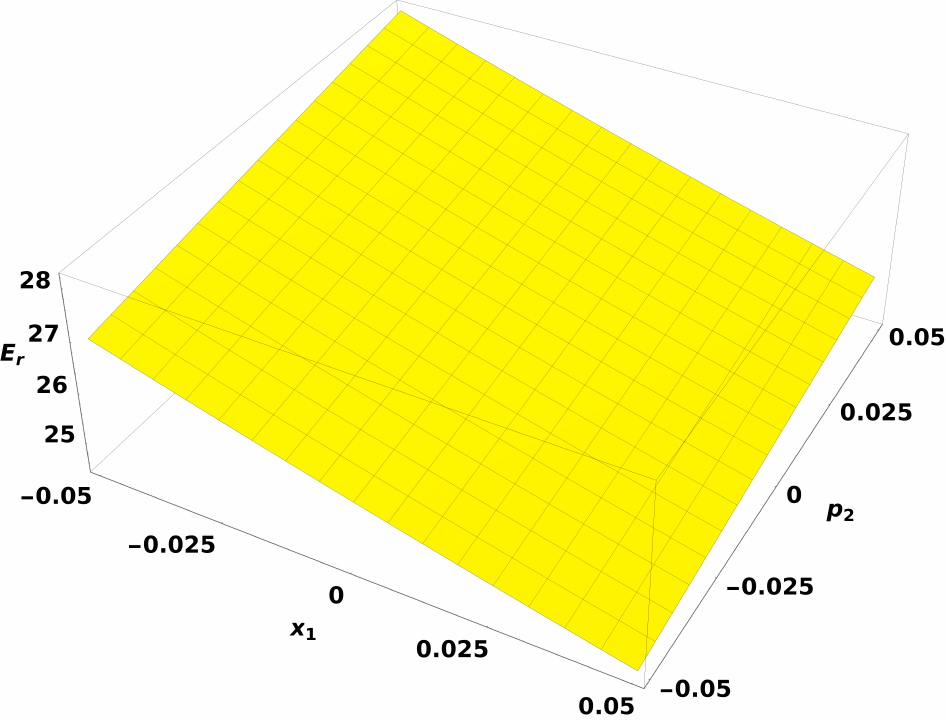}}
			\captionsetup{width=\linewidth, aboveskip=4pt}
			\caption{
				Real eigenvalue plot of reality of spectrum for the second root of $\beta_3$ near origin, where mass profile is 
				$m_r = cx_1 - dp_2 + e_1$ and
				$m_i = dx_1 + cp_2 + e_2$ in the extended complex plane.
			}
			\label{fig9}
		\end{minipage}
		\hfill
		\begin{minipage}[b]{0.47\textwidth}
			\centering
			\fbox{\includegraphics[scale=0.37]{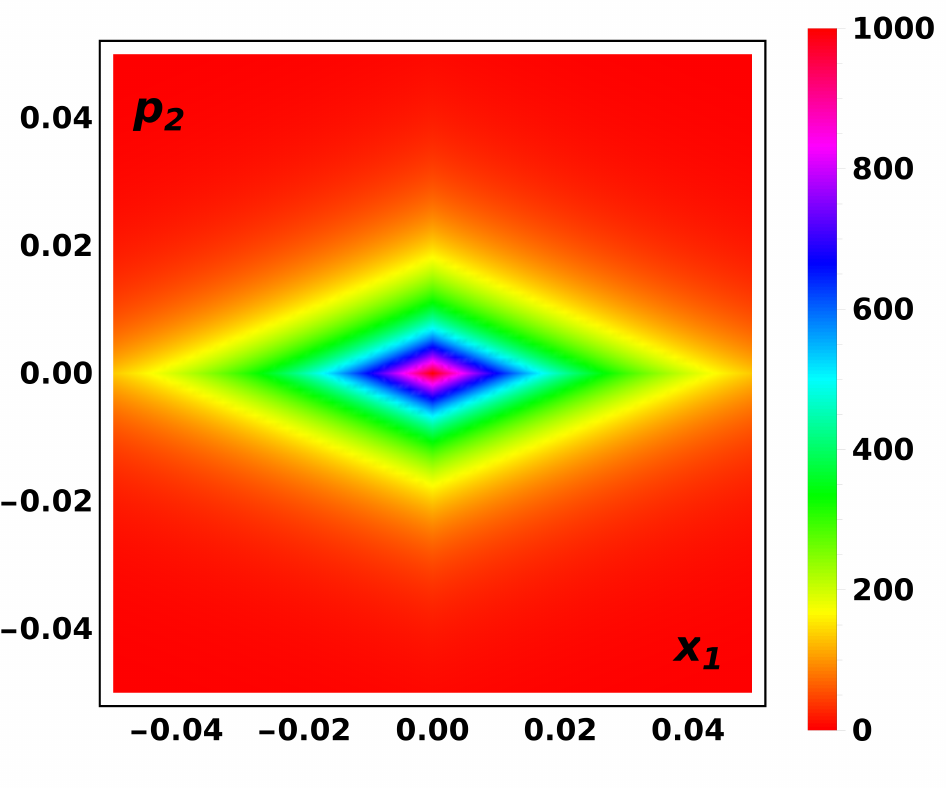}}
			\captionsetup{width=\linewidth, aboveskip=4pt}
			\caption{
				Probablity density plot of reality of spectrum for the second root of $\beta_3$ near origin, where mass profile is 
				$m_r = cx_1 - dp_2 + e_1$ and
				$m_i = dx_1 + cp_2 + e_2$ in the extended complex plane.
			}
			\label{fig10}
		\end{minipage}
	\end{figure}

	While real eigenvalues and probability density functions can be evaluated for both roots within the region representing the existence of a diatomic molecule, it is important to highlight that the normalization conditions given by Equations \ref{eq32} and \ref{eq33} are satisfied only in the purple-shaded regions of the plots, and not in the pink regions. Notably, the origin in figure \ref{fig8b} lies within the purple region, whereas in figure \ref{fig8a}, it is located in the pink region. This clearly indicates that only the second root of $\beta_3$ yields a properly normalized eigenfunction. The behavior of the corresponding real eigenvalue and probability current density for this valid root is depicted in figures \ref{fig9} and \ref{fig10}.
	
	To analyze the behaviour of Complex Morse potential admitting real eigenvalues, we insert the conditions imposed on the mass profiles in Cases I(a) and II(a) from the previous section. 
	
	\subsection{Case I(b), $\frac{dm_r}{dx_1} = 0$}
	
	The general expression of roots of $\beta_3$ is written in the form
	\begin{equation}
		\beta_3 = \frac{m^2 \lambda_1 \mp \gamma_1}{2 m^2 \lambda_2}, \label{eq49}
	\end{equation}
	whereas, the value of $\alpha_1$ and $\beta_1$ parameters take the value,
	
	\begin{eqnarray}
		\alpha_1 = \frac{|a|^2m^2(m_i a_r - m_r a_i)-\lambda_2 m_i m'_i \mp a_i \gamma_1}{2 m^2 \lambda_2}, \label{eq50} \\
		\beta_1 = \frac{|a^2|m^2(m_i a_i - m_r a_r)+\lambda_2 m_r m'_i \mp a_r \gamma_1}{2 m^2 \lambda_2}. \label{eq51}
	\end{eqnarray}
	
\begin{figure*}[h!]
		\centering
		\begin{subfigure}[b]{0.4\paperwidth}
			\centering
			\fbox{\includegraphics[scale=0.32]{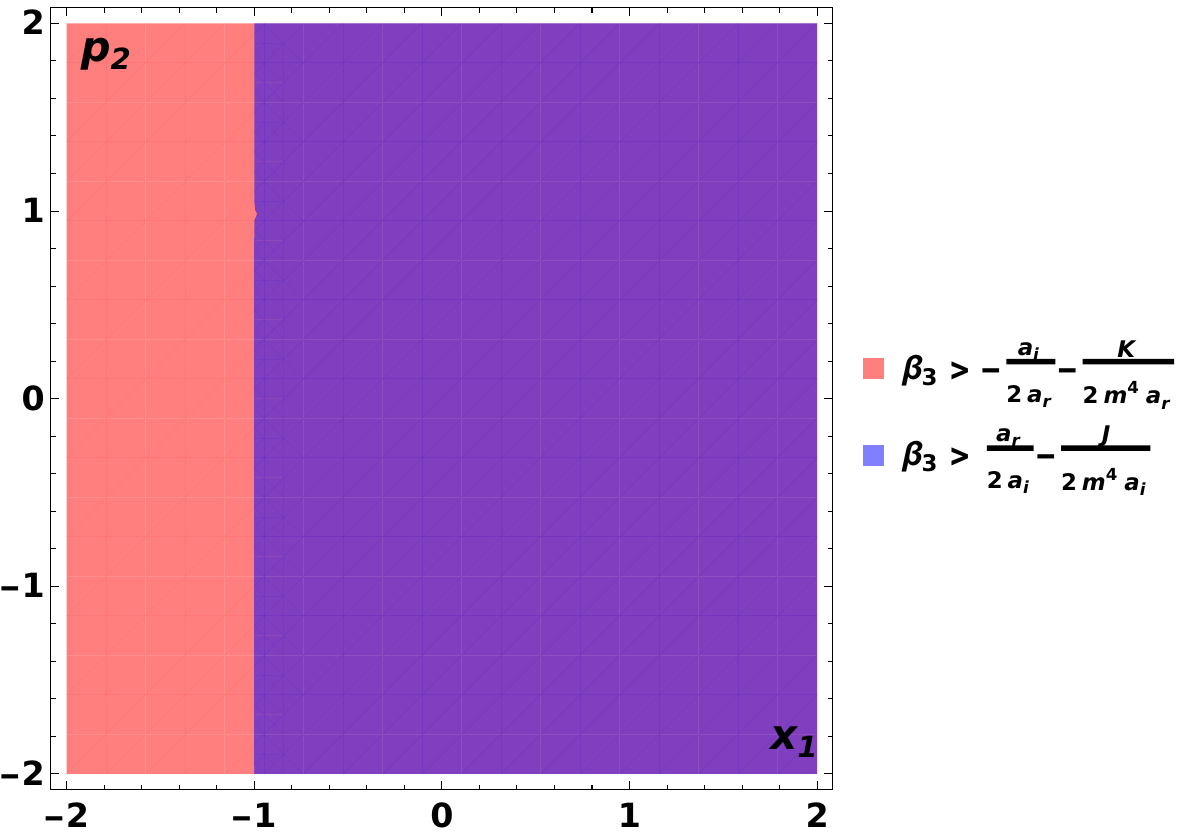}}
			\caption{}
			\label{fig11a}
		\end{subfigure}
		\hspace*{-5em}
		\begin{subfigure}[b]{0.4\paperwidth}
			\centering
			\fbox{\includegraphics[scale=0.32]{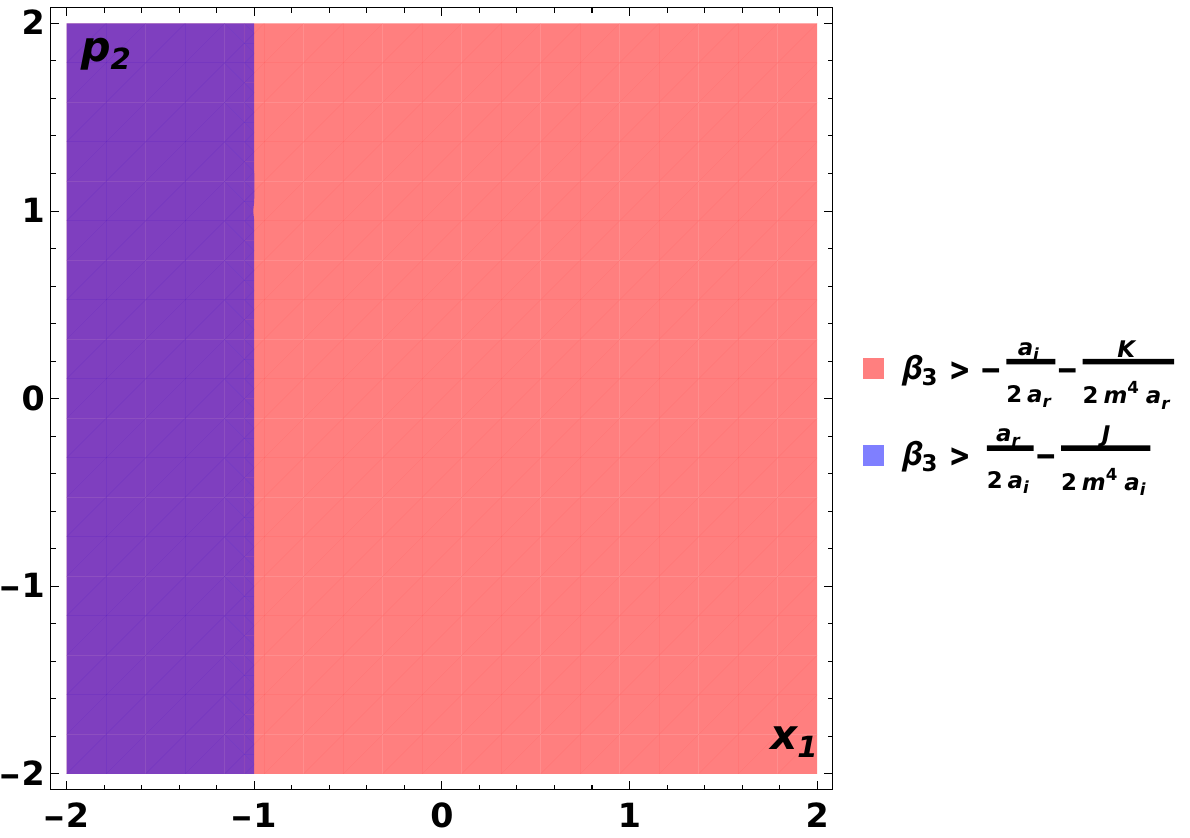}}
			\caption{}
			\label{fig11b}
		\end{subfigure}
		\captionsetup{aboveskip=4pt}
		\caption{Normalization condition plot of reality of spectrum for Case I(b) in extended complex plane associated with (a) first root of $\beta_3$ (b) second root of $\beta_3$.}
		\label{fig11} 	 
	\end{figure*}

\begin{figure}[h!]
		\centering
		\begin{minipage}[b]{0.47\textwidth}
			\centering
			\fbox{\includegraphics[scale=0.4]{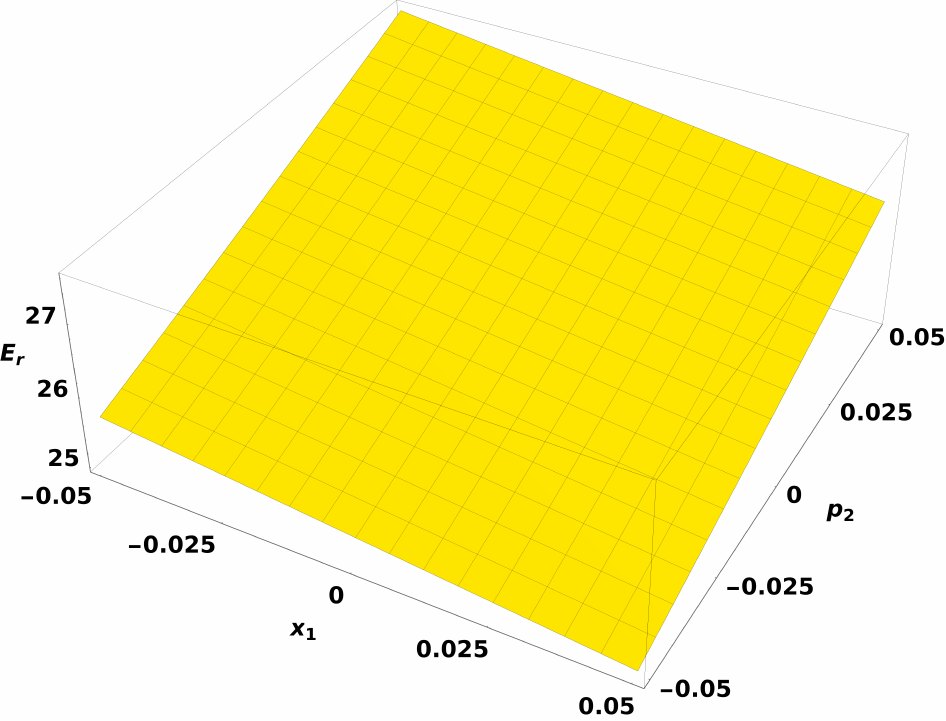}}
			\captionsetup{width=\linewidth, aboveskip=4pt}
			\caption{
				Real eigenvalue plot of Case I(b) for the first root of $\beta_3$ near origin in the extended complex plane.
			}
			\label{fig12}
		\end{minipage}
		\hfill
		\begin{minipage}[b]{0.47\textwidth}
			\centering
			\fbox{\includegraphics[scale=0.37]{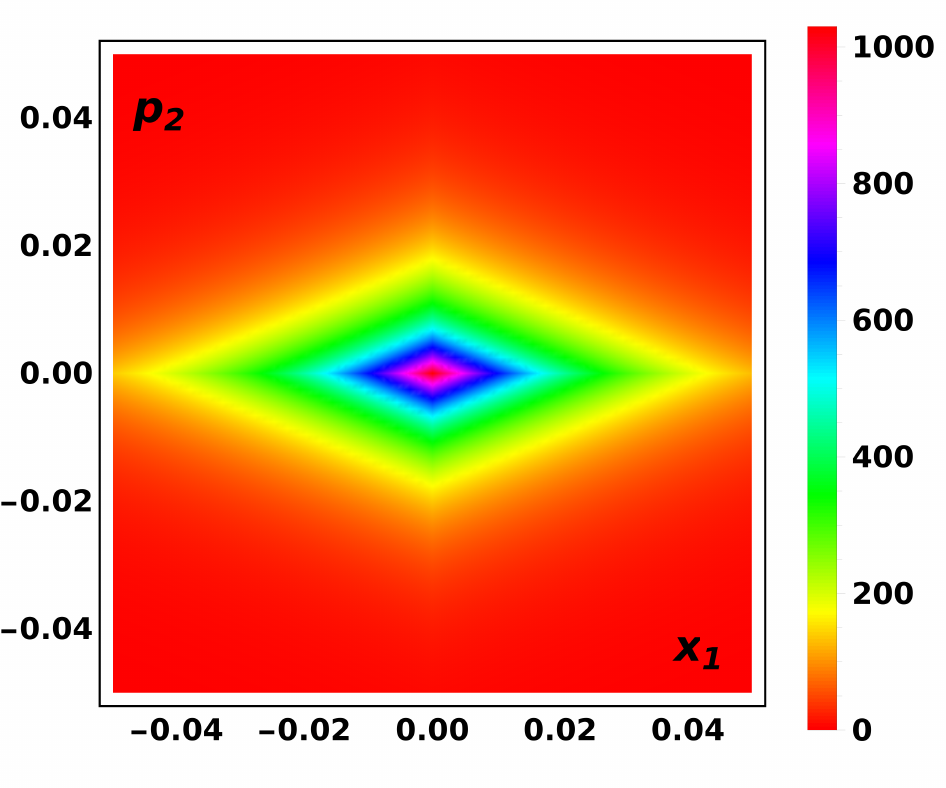}}
			\captionsetup{width=\linewidth, aboveskip=4pt}
			\caption{
				Probablity density plot of Case I(b) for the first root of $\beta_3$ near origin in the extended complex plane.
			}
			\label{fig13}
		\end{minipage}
	\end{figure}

	Further, the real eigenvalue is obtained in the form, 
	\begin{equation}
		E_r = \frac{(m_i^2 m_r |a|^4 + m^2 a_i a_r \lambda_3){m'_i}^2+|a|^4m^4 \lambda_1 \mp m^2 \gamma_1}{4 m^4 \lambda_2^2} \label{eq52}
	\end{equation}
	where, the variables $\lambda_1$, $\lambda_2$, $\lambda_3$ and $\gamma_1$ are functions of masses and dissociation parameters $a$ defined as,
\begin{subequations}
	\begin{align}
	\lambda_1 &= m_r (a_r^2 - a_i^2) + m_i(2 a_i a_r), \label{eq53a} \\
	\lambda_2 &= m_i (a_i^2 - a_r^2) + m_r(2 a_i a_r), \label{eq53b} \\
	\lambda_3 &= m_i (a_i^2 - a_r^2) - m_r(2 a_i a_r). \label{eq53c} 
	\end{align}
\end{subequations}
		
	\begin{equation}
		\gamma_1 = \sqrt{|a|^4 m^6 + \lambda_2(m_i^2-3m_r^2) m_i m'_i}. \label{eq54}
	\end{equation}
	Figures \ref{fig11a} and \ref{fig11b} present the plots of the normalization conditions to identify the permissible values of $\beta_3$ for this scenario. It is clearly observed that the first root of $\beta_3$ corresponds to a region near the origin where the eigenfunction is properly normalized. Additionally, the behavior of the real eigenvalue and the associated probability density is illustrated in figures \ref{fig12} and \ref{fig13}, providing further insight into the physical characteristics of the system.
	
	\subsection{Case II(b), $\frac{dm_i}{dx_1} = 0$} 
	
\begin{figure*}[h!]
		\centering
		\begin{subfigure}[b]{0.4\paperwidth}
			\centering
			\fbox{\includegraphics[scale=0.32]{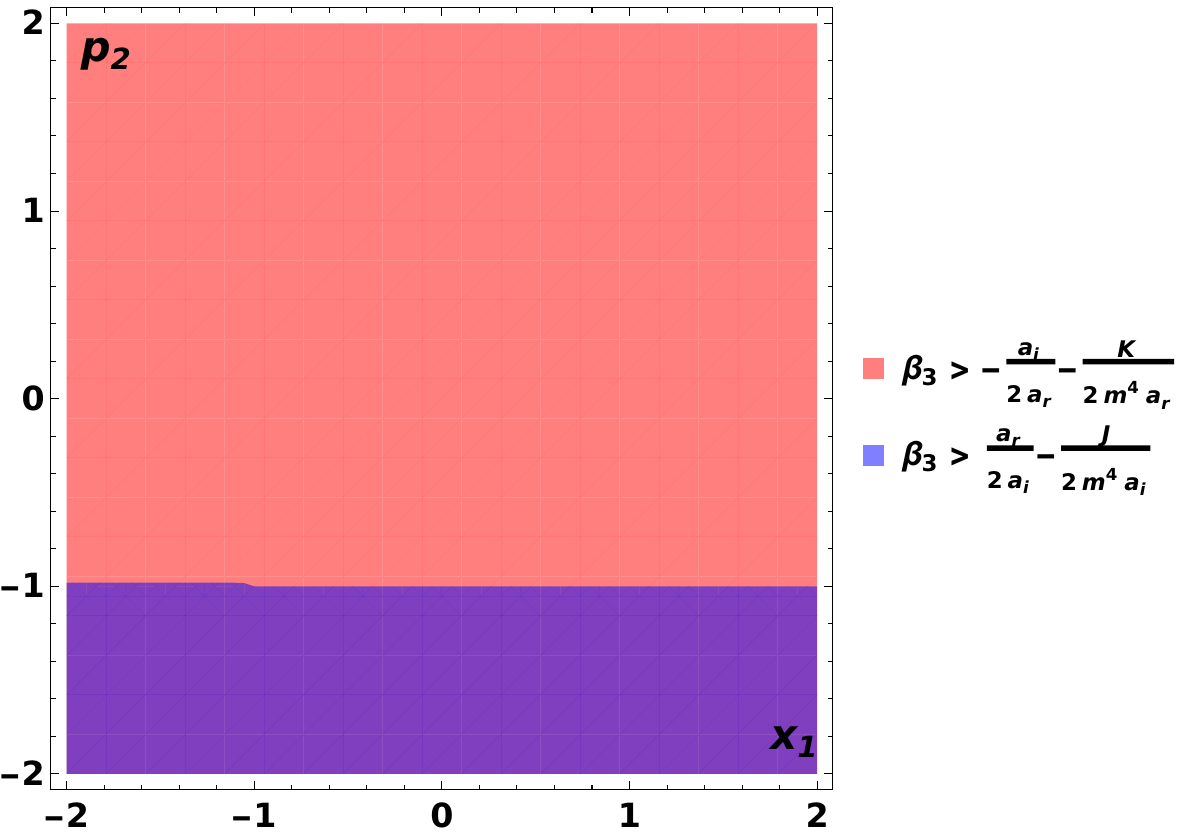}}
			\caption{}
			\label{fig14a}
		\end{subfigure}
		\hspace*{-5em}
		\begin{subfigure}[b]{0.4\paperwidth}
			\centering
			\fbox{\includegraphics[scale=0.32]{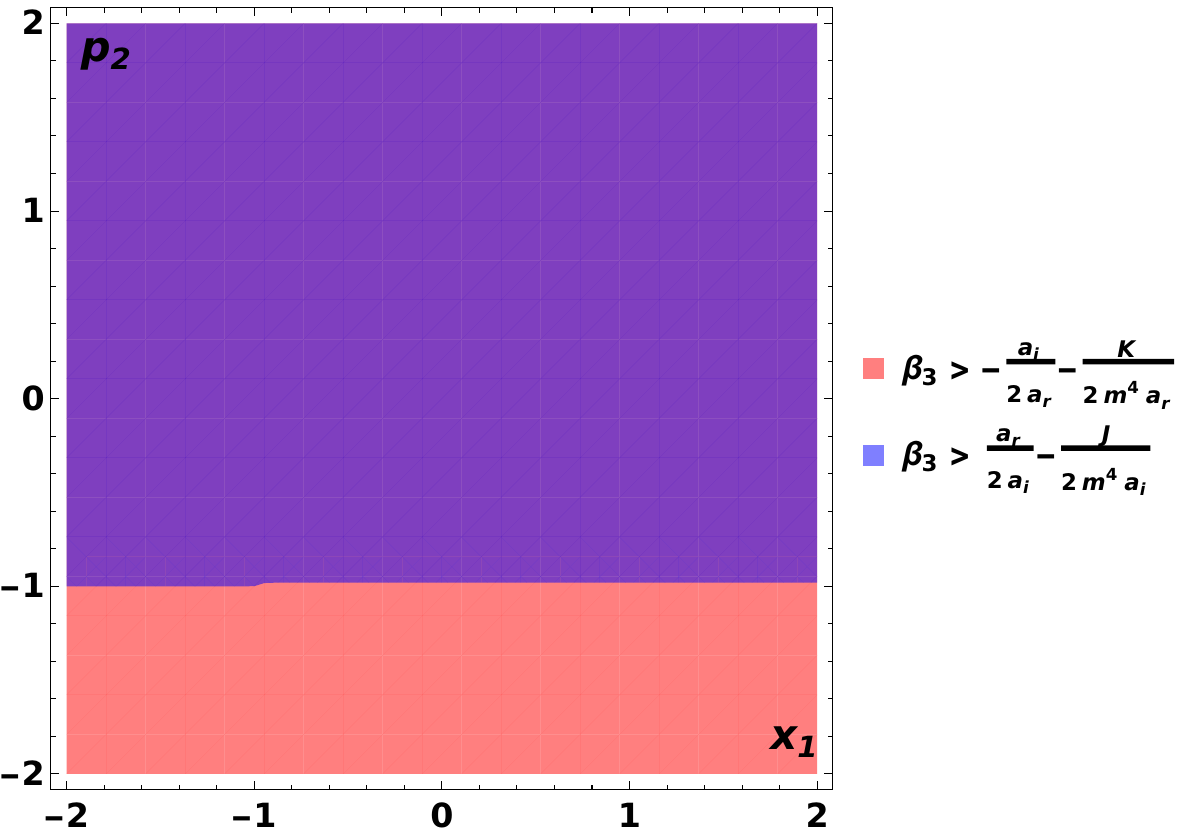}}
			\caption{}
			\label{fig14b}
		\end{subfigure}
		\captionsetup{aboveskip=4pt}
		\caption{Normalization condition plot of reality of spectrum for Case II(b) in extended complex plane associated with (a) first root of $\beta_3$ (b) second root of $\beta_3$.}
		\label{fig14} 	 
	\end{figure*}

\begin{figure}[h!]
		\centering
		\begin{minipage}[b]{0.47\textwidth}
			\centering
			\fbox{\includegraphics[scale=0.4]{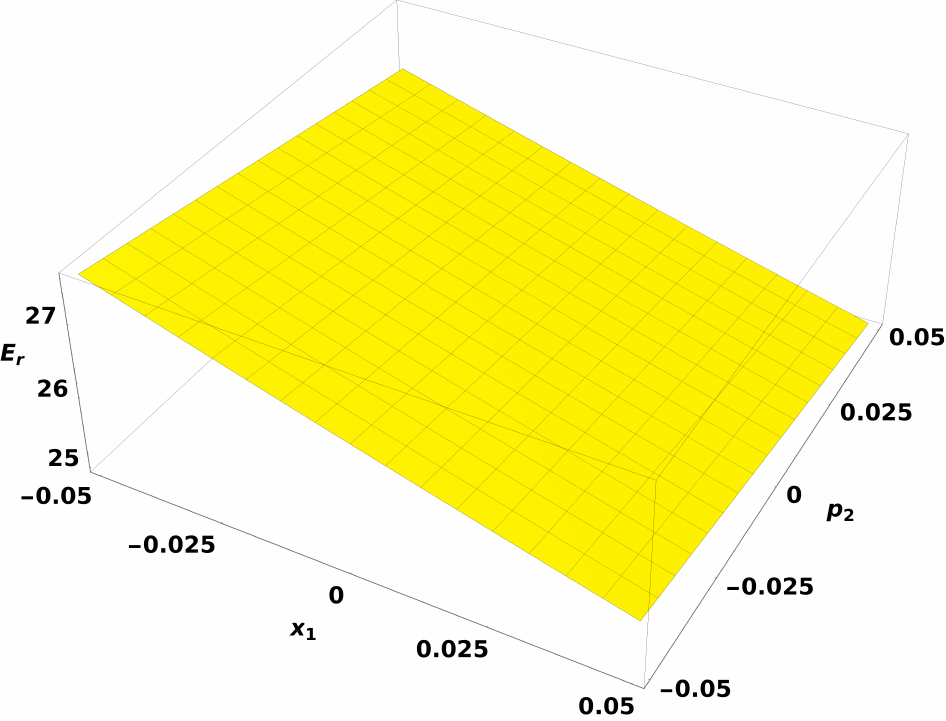}}
			\captionsetup{width=\linewidth, aboveskip=4pt}
			\caption{
				Real eigenvalue plot of Case II(b) for the second root of $\beta_3$ near origin  in the extended complex plane.
			}
			\label{fig15}
		\end{minipage}
		\hfill
		\begin{minipage}[b]{0.47\textwidth}
			\centering
			\fbox{\includegraphics[scale=0.37]{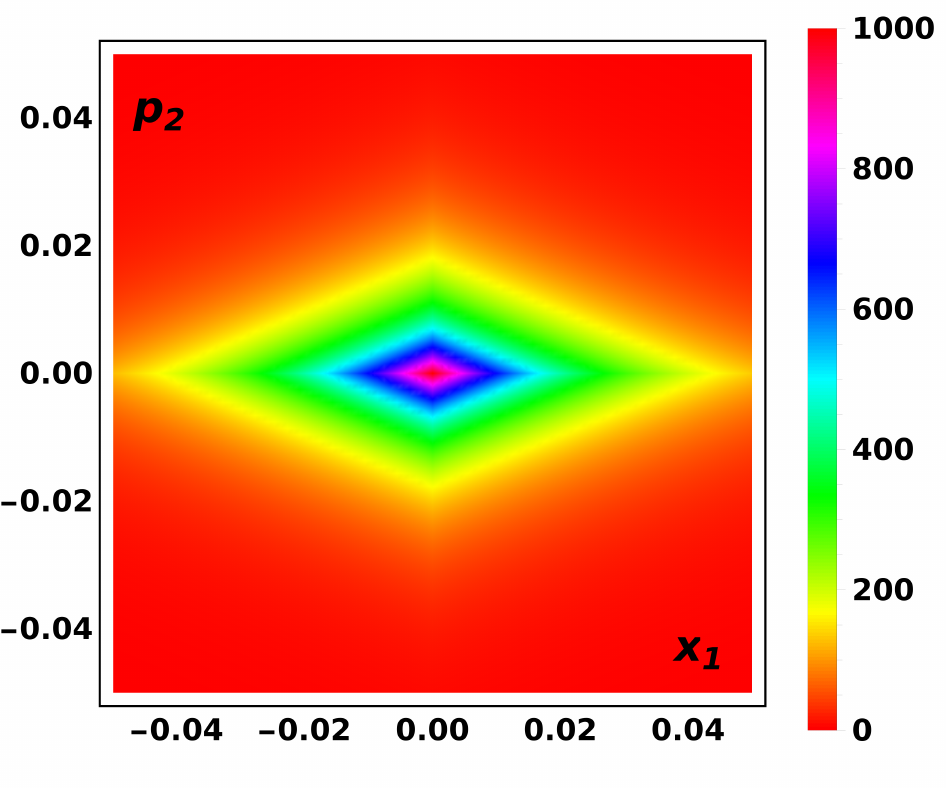}}
			\captionsetup{width=\linewidth, aboveskip=4pt}
			\caption{
				Probablity density plot of Case II(b) for the second root of $\beta_3$ near origin  in the extended complex plane.
			}
			\label{fig16}
		\end{minipage}
	\end{figure}

	Similar to above case, the value of parameters and eigenvalue is deduced as, 
	
	\begin{align}
		\beta_3 &= \frac{m^2 \lambda_1 \mp \gamma_2}{2 m^2 \lambda_2}, \label{eq55} \\	
		\alpha_1 &= \frac{|a|^2m^2(m_i a_r - m_r a_i)-\lambda_2 m_r m'_r \mp a_i \gamma_2}{2 m^2 \lambda_2}, \label{eq56} \\
		\beta_1 &= \frac{|a^2|m^2(m_i a_i + m_r a_r)-\lambda_2 m_i m'_r \mp a_r \gamma_2}{2 m^2 \lambda_2}, \label{eq57} \\
		E_r &= \frac{-(m_i^2 m_r |a|^4 + m^2 a_i a_r \lambda_3){m'_r}^2+|a|^4m^4 \lambda_1 \mp m^2 \gamma_2}{4 m^4 \lambda_2^2}. \label{eq58}
	\end{align}
	Where all the $\lambda$ parameters are the same as enumerated in equations \ref{eq53a}-\ref{eq53b} while $\gamma_2$ will be written as
	\begin{equation}
		\gamma_2 = \sqrt{|a|^4 m^6 - \lambda_2(m_i^2-3m_r^2)m_i m'_r}. \label{eq59}
	\end{equation}
	The normalization condition plots shown in figures \ref{fig14a} and \ref{fig14b} clearly indicate that the second root of $\beta_3$ satisfies the normalization criteria in the region near the origin. Correspondingly, figures \ref{fig15} and \ref{fig16} depict the eigenvalue and probability density within this region, illustrating the behavior of the system for this case where the energy spectrum remains real. 
	
	\section{Conclusion} \label{sec7}
	
	In this study, we have analyzed the complex Morse potential with position-dependent mass (PDM) by solving the corresponding position-dependent mass Schrödinger equation (PDMSE). We derived the eigenvalues and eigenfunctions and addressed the critical aspect of normalization. Due to the non-Hermitian nature of complex potentials and the variable mass distribution, the eigenfunctions are generally not square-integrable in the traditional sense. Therefore, we adopted a modified normalization condition defined as a two-dimensional integral over the phase space:
	
	$$\int_{-\infty}^{\infty} \int_{-\infty}^{\infty} \psi^*\left(x_1,p_2\right) \psi\left(x_1,p_2\right) dx_1 dp_2 = \hbox{constant}.$$
	
	The convergence of this integral is essential for maintaining the probabilistic interpretation of quantum mechanics. A key achievement of the present study was the formulation of an appropriate normalization condition suited to complex systems with PDM. Using this criterion, the normalized eigenfunction was obtained (as shown in Equation \ref{eq34}), with the normalization constant evaluated in Equation \ref{eq31} under the constraints specified in Equations \ref{eq32} and \ref{eq33}. This ensures the physical validity of the wavefunctions, enabling stable, interpretable, and physically meaningful expectation values.
	
	Our analysis demonstrates that, under suitable parameter constraints, the Complex Morse Potential with PDM can support real eigenvalues—an essential requirement for observable physical quantities. The probability density plots (figures \ref{fig3a}, \ref{fig7a} and \ref{fig7b}) exhibit finite peaks in phase space, indicating that the bound states are localized, and particles remain confined to specific regions. These results confirm that the system supports stable, localized bound states and preserves many qualitative features of the real Morse potential.
	
	Furthermore, our findings underscore a significant point: the reality of the energy spectrum is more fundamental than the Hermiticity of the Hamiltonian. The complex Morse system with PDM can exhibit real eigenspectra due to underlying symmetries or parameter conditions, even when the Hamiltonian itself is non-Hermitian. Complex potentials with PDM are analogous to refractive index profiles in PT-symmetric optical systems. Our findings could inform the design of optical lattices or waveguides. Since many open systems are inherently non-Hermitian, the framework developed here can model gain-loss systems, decoherence, or energy dissipation \cite{ruter2010observation, musslimani2008optical, De2019con, el2018non, peng2025observation}. Finally, we speculate that such systems may have deeper implications in high-energy or cosmological contexts \cite{sarathi2021application, borzou2021estimation}. The complex Morse potential with position-dependent mass may provide a theoretical framework for describing exotic quantum systems such as dark matter. Given that dark matter constitutes a significant portion of the Universe, our model suggests that bound states governed by complex PDM potentials could offer insight into its quantum behaviour. These findings also raise the possibility that visible matter emerged via phase transitions or symmetry breaking, leading to effective real Morse-like potentials in the post-early-Universe evolution.
	
	The methods and insights presented here contribute meaningfully to the growing body of literature on non-Hermitian quantum mechanics and provide a viable quantum model for systems with position-dependent complex mass, a concept that could play crucial role in both fundamental physics and applied quantum physics.

	\end{document}